%% file: main.tex
\documentclass[12pt]{article}
\usepackage[utf8]{inputenc}
\usepackage{times}          
\usepackage{textgreek}      
\usepackage{bm}             

\usepackage[margin=1in]{geometry}
\usepackage{setspace}       
\usepackage{titlesec}      
\usepackage{lineno}         

\usepackage{amsmath, amssymb}
\usepackage{siunitx}        
\newcommand{\asym}[3]{\ensuremath{#1^{+#2}_{-#3}}}
\usepackage{graphicx}
\usepackage{booktabs}       
\usepackage{caption}        
\usepackage{subcaption}     
\usepackage{multirow}
\usepackage{threeparttable} 

\usepackage{url}
\usepackage[hidelinks]{hyperref} 
\usepackage[
    style=nature,    
    backend=biber,   
    sorting=none,     
    defernumbers=true,
    eprint=false
]{biblatex}
\DeclareFieldFormat[inproceedings]{title}{#1}
\DeclareCaptionType{edfigure}[Extended Data Figure][List of Extended Data Figures]
\DeclareCaptionType{sfigure}[Supplementary Figure][List of Supplementary Figures]
\DeclareCaptionType{stable}[Supplementary Table][List of Supplementary Tables]

\newcommand{\bOmega}{\bm{\Omega}}
\titleformat{\section}{\bfseries\large}{\thesection}{1em}{}
\titleformat{\subsection}{\bfseries}{\thesubsection}{1em}{}

\title{\bfseries High-energy neutrino emission from the Milky Way}

\author{
The IceCube Collaboration\thanks{A list of authors and their affiliations appears at the end of the paper.}
}

\date{}

\begin{document}

\maketitle

\vspace{1em}

\textbf{
The Milky Way hosts astrophysical objects that accelerate cosmic rays to energies beyond the reach of terrestrial particle accelerators. It remains a longstanding goal to locate the sites of these powerful Galactic engines and understand how cosmic rays propagate through the Galaxy, leading to the production of high-energy neutrinos. In this paper, we combine event morphologies characteristic of all three neutrino flavours and apply recent improvements in ice modelling, calibration and reconstruction to 12 years of IceCube data. With a predefined, global analysis we establish high-energy neutrino emission from the Galactic plane at \SI{5.7}{\textsigma} significance. A further study shows that the inner region of the Galaxy is a prominent neutrino source, with \num{217} shower events with visible energy above \SI{5}{\tera\eV} compared with an expected background of \num{154.4 \pm 4.1}. These results herald a new era of Galactic multi-messenger astronomy, creating new opportunities to study cosmic-ray propagation and probe neutrino properties over kiloparsec distances.}

\section*{Main}

A decade ago IceCube discovered a diffuse flux of high-energy neutrinos reaching us from the Universe~\autocite{IceCube:2013low}. At the time, the flux appeared isotropic, with no resolved contribution from the Milky Way. In contrast to electromagnetic observations, where the Milky Way is a prominent source in the sky, the total neutrino emission of distant extragalactic sources outshines our Galaxy by one order of magnitude. Previous searches have looked for Galactic neutrino emission~\autocite{IceCube:2017trr,ANTARES:2018nyb,IceCube:2019lzm,IceCube:2025zyb}, and IceCube has reported evidence of a high-energy neutrino flux from the Galactic plane~\autocite{IceCube:2023ame} (GP). Here we report its observation at \SI{5.7}{\textsigma} post-trial significance, establishing the Milky Way as the first astrophysical source of high-energy neutrinos to exceed the \SI{5}{\textsigma} discovery threshold. The observed emission is concentrated toward the inner region of our Galaxy and is compared against different models of high-energy hadronic emission and Galactic cosmic-ray transport.

Although the high-energy astrophysical neutrino flux is largely extragalactic~\autocite{IceCube:2013low}, a Galactic component is produced when cosmic rays accelerated within the Milky Way interact with interstellar gas. Galactic CRs constitute a non-thermal component of the Milky Way with an energy density comparable to that of magnetic fields and starlight, and their interactions with the interstellar gas therefore represent a channel of energy dissipation within the interstellar medium~\autocite{Ferriere:2001rg}. A diffuse Galactic flux of high-energy neutrinos is produced through decays of secondary particles generated in the CR collisions.
Thus, neutrinos trace collisions of CRs with their ambient environment, and preserve information of their origin, even when those regions are inaccessible to electromagnetic observations.

Neutrinos produced in the Milky Way are expected to reach Earth as a mixture of $\nu_e$, $\nu_\mu$ and $\nu_\tau$ due to the quantum phenomenon of flavour oscillations~\autocite{ParticleDataGroup:2024cfk}, motivating an analysis sensitive to all three flavours. Because neutrinos only interact weakly and the fluxes of extraterrestrial neutrinos at TeV energies and above are low (on the order of \qty{1}{\per\km\squared\per\minute} above \qty{100}{\tera \eV}), a large-volume detector is needed for significant detection rates in this energy regime. The IceCube Neutrino Observatory instruments a cubic-kilometre of South Pole ice with \num{5160} single-photon sensitive digital optical modules (DOMs) that detect the Cherenkov radiation produced by outgoing, charged particles from neutrino interactions~\autocite{IceCube:2016zyt}. The ice thus serves as both the target and detection medium. Dominant backgrounds in searches for astrophysical neutrinos are downward-going muons (from the southern sky) and neutrinos~\autocite{Abbasi:2021qfz,IceCube:2022der} produced by CR interactions in the atmosphere. Atmospheric muons can be suppressed by selecting events with interaction vertices inside or near the instrumented volume~\autocite{IceCube:2023ame,IceCube:2025zyb}.

Based on their light deposition patterns, neutrino interactions in IceCube are usually classified as track-like or shower-like (also sometimes called cascade) events, with tracks dominated by charged-current $\nu_\mu$ interactions and showers by charged-current $\nu_e$ and $\nu_\tau$ interactions as well as neutral-current interactions. At \(E_{\bar{\nu}_e}\simeq \SI{6.3}{\peta \eV}\), electron antineutrinos can also resonantly scatter off atomic electrons in the ice to produce an on-shell \(W^-\) boson~\autocite{Glashow:1960zz,IceCube:2021rpz}. Track-like events can be further subdivided into starting-tracks, where the neutrino is reconstructed to have interacted within the instrumented region of the detector, and through-going tracks, where the neutrino interaction vertex is thought to be outside the instrumented region and a resulting muon is detected as it traverses through. Physics parameters of interest, such as the arrival direction (right ascension, $\hat{\alpha}$, and declination, $\hat{\delta}$), angular uncertainty, $\hat{\sigma}$, and energy proxy, $\hat{E}$, used in the likelihood analysis described in this article, can be estimated using reconstruction routines. The hat notation is used throughout the text to denote a reconstructed quantity.

Combining event selections targeting different morphologies, each contributing additional statistics with different signal-to-background ratios, improves sensitivity beyond that of the individual components. Such gains have previously been demonstrated in IceCube measurements of isotropic astrophysical neutrino emission~\autocite{IceCube:2015gsk,IceCube:2025tgp} and in point-source searches~\autocite{IceCube:2025lev}. Joint ANTARES-IceCube analyses have also searched for Galactic-plane emission and extended Galactic sources using all-flavour information~\autocite{ANTARES:2018nyb,ANTARES:2020srt}. Here, we combine three complementary event selections in a unified all-flavour analysis of diffuse Galactic neutrino emission~\autocite{IceCube:2023gtp,Thiesmeyer:2025qgo}. Unlike previous IceCube Galactic-plane searches, this framework simultaneously incorporates shower-dominated, starting-track, and through-going-track samples within a single likelihood analysis.

The first sample is shower-dominated;
it builds on the previous 10-year IceCube Galactic-plane analysis~\autocite{IceCube:2023ame} by extending the exposure to 12 years and incorporating updated detector calibration, simulation, and reconstruction. As models of Galactic neutrino emission predict spatially extended sources, the larger directional uncertainty of shower-like events 
can be compensated by statistics. The second sample consists of starting tracks~\autocite{IceCube:2025zyb}, which target the full sky with better angular resolution but lower statistics than showers. The third is a northern-track sample~\autocite{Abbasi:2021qfz,IceCube:2022der} ($\hat{\delta} > \ang{-5}$), dominated by through-going muons that are produced outside the detector and traverse through the instrumented region. Because the structure scale of model predictions, or ``templates'', is comparable to the angular resolution of shower events, and concentrated in the southern hemisphere where low-energy track events are harder to distinguish from CR background, our sensitivity comes primarily from the shower subsample.
The expected improvement in \SI{5}{\textsigma} discovery potential (the median
flux required for a \SI{5}{\textsigma} detection) from extending the shower sample
from 10 to 12 years is $\sim$8\% for the \textit{Fermi}-LAT $\pi^0$ model~\autocite{Fermi-LAT:2012edv}, with
the two track samples contributing a further $\sim$17\%. Values for other tested
templates are given in Supplement, Tab.1.

\begin{figure}[ht!]

    \begin{center}
    \begin{subfigure}[t]{0.44\linewidth}
        \includegraphics[width=\linewidth]{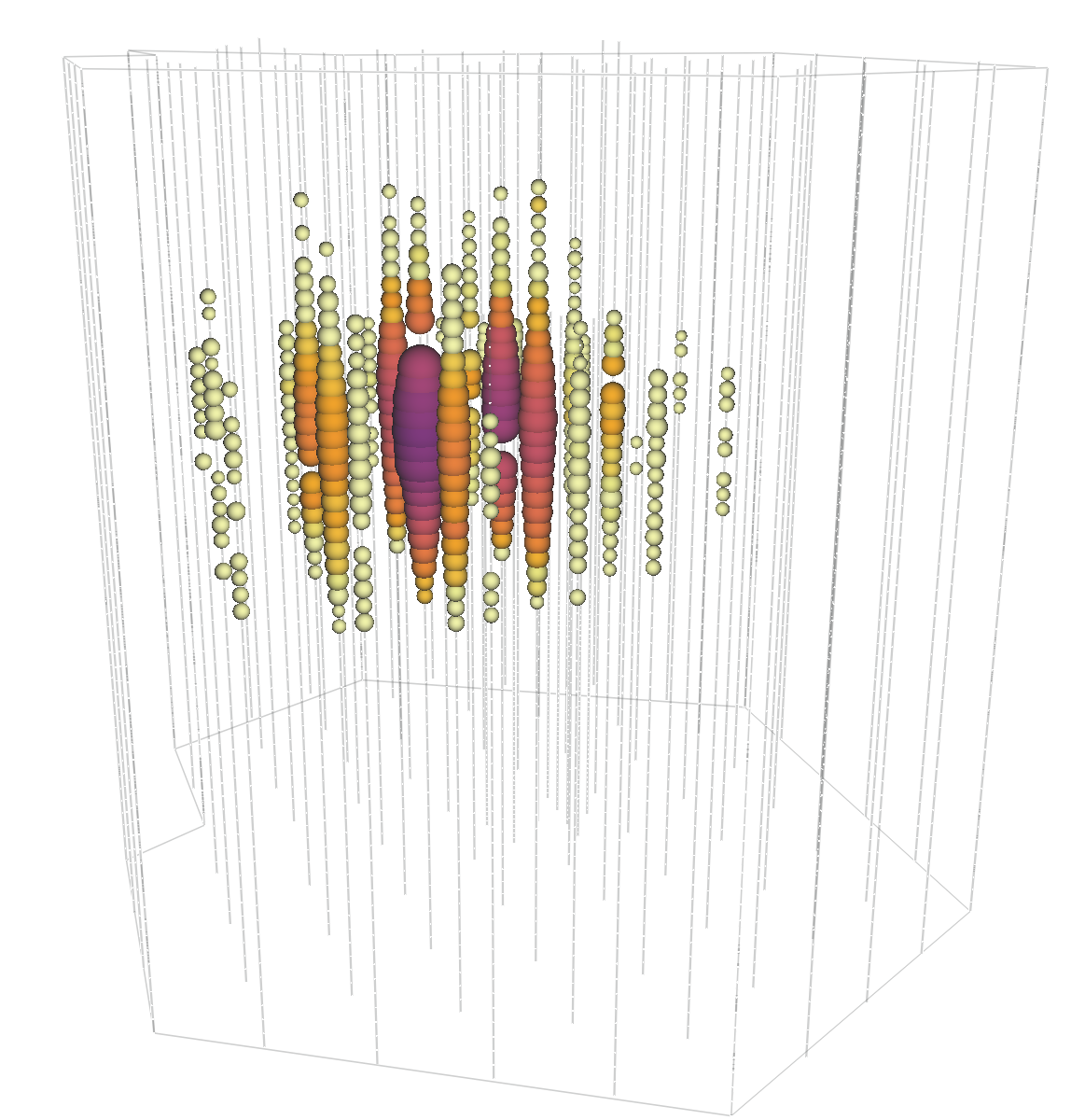}
        \caption{}
    \end{subfigure}
    \begin{subfigure}[t]{0.44\linewidth}
        \includegraphics[width=\linewidth]{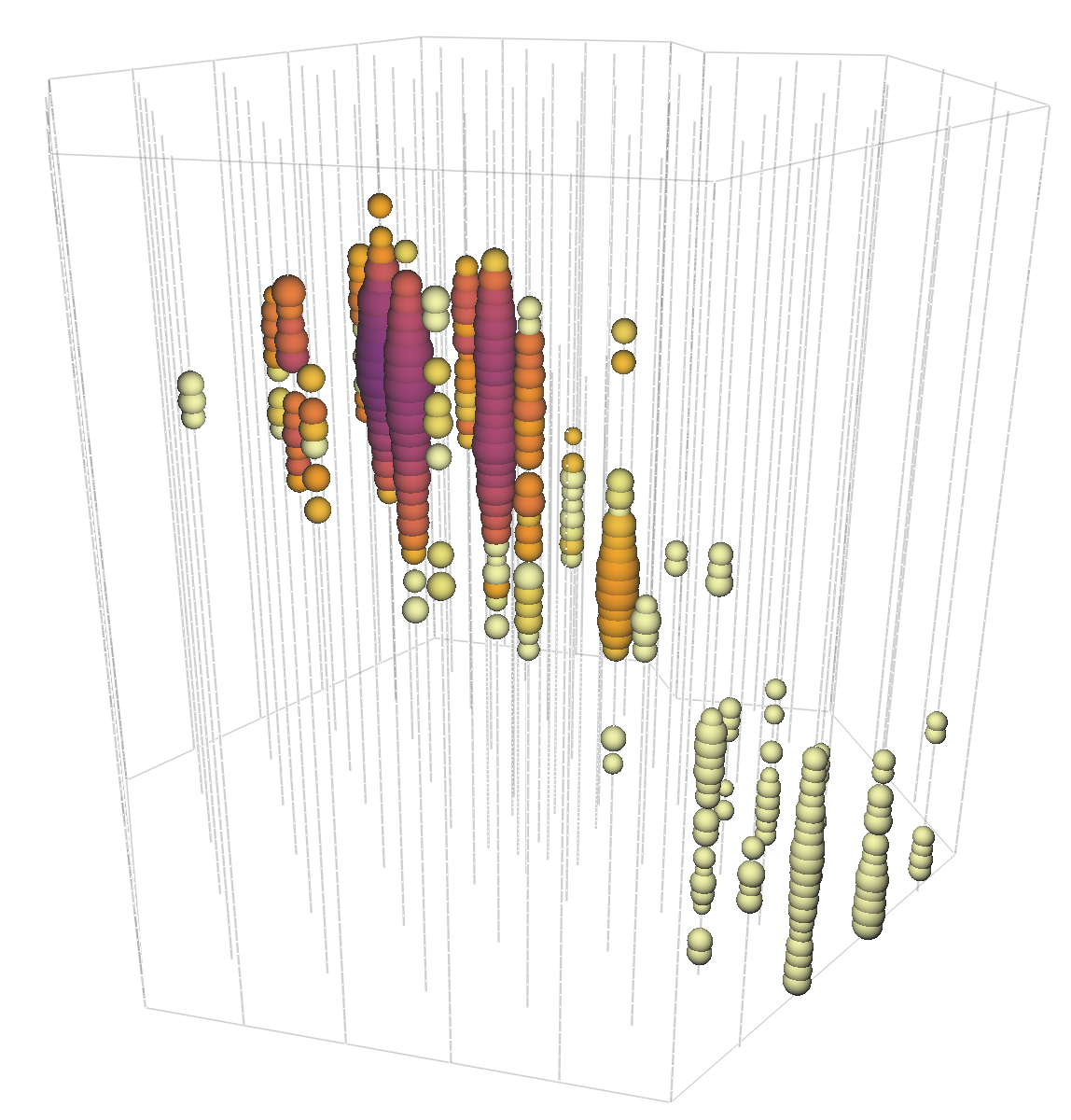}
        \caption{}
    \end{subfigure}
    \end{center}

    \caption{\textbf{Two example events.}
    We show the two neutrino events that individually yield the largest contributions to the test statistic (TS) in this analysis. Each coloured sphere corresponds to signal in a digital optical module (DOM), with size corresponding to the total charge and colour indicating the time of the first detected photon. Magenta (yellow) corresponds to earlier (later) arrival times.
    \textbf{(a)} Shower event with a reconstructed energy of $\sim$\SI{1000}{\tera \eV}~\autocite{IceCube:2013cdw,IceCube:2013low}, consistent with a $\nu_e$ charged--current interaction or a neutral--current interaction of any flavour. 
    \textbf{(b)} Track event with an interaction vertex contained within the detector, likely originating from a $\nu_\mu$ charged--current interaction.
    }

    \label{fig:eventview}
\end{figure}

The data are composed of \num{1069454} events, of which \num{984255} (\num{85199}) are classified as tracks (showers). Figure~\ref{fig:eventview} shows event displays of the shower-like (left panel) and track-like (right panel) events that provide the strongest support for a Galactic signal. See the Methods section for a detailed description of the datasets.

Event reconstructions for the different channels are optimized for their respective morphologies. In particular, the meaning of $\hat{E}$ can differ depending on the sample; it is a proxy of the total visible energy from Cherenkov radiation for showers~\autocite{IceCube:2024csv}, and the muon energy at the detector entrance or the neutrino energy for through-going~\autocite{IceCube:2022der} or starting~\autocite{IceCube:2024fxo} tracks, respectively. For shower-like events, photon propagation in the ice is a leading source of directional uncertainty. With respect to the previous analysis~\autocite{IceCube:2023ame}, the Monte Carlo (MC) simulation and reconstruction used for the shower sample now include effects due to ice birefringence~\autocite{tc-18-75-2024,IceCube:2024csv}. Microscopic birefringence of ice crystals accumulates to a macroscopic axial asymmetry in light propagation aligned with the glacial-flow direction; neglecting this effect previously introduced azimuth-dependent reconstruction biases, as illustrated in Extended Data Fig.~\ref{fig:iceman_azimuth}. The updates also include improved descriptions of ice-layer undulations~\autocite{IceCube:2023qua} and the refrozen ice within the drill hole~\autocite{IceCube:2013llx}. Tests using MC indicate a reduction in systematic biases in the reconstructed zenith angle compared to prior work, and that the median shower angular resolution improves over the sensitive energy range by roughly a factor of 1.5--2 as shown in Extended Data Fig.~\ref{fig:iceman_reco_sim}. These improvements increase confidence in the accuracy and robustness of the result.

We tested correlations between reconstructed event directions and diffuse Galactic emission using an unbinned maximum likelihood that is constructed from a mixture of signal and background probability density functions (PDF)~\autocite{Braun:2009wp}. The null hypothesis assumes arrival directions uniform in right ascension, while the alternative hypothesis includes a Galactic signal described by one of four templates. The likelihood incorporates both directional and energy information for each event. The test statistic (TS) is defined as the logarithm of the ratio between the maximum likelihood of the alternative hypothesis to that of the null hypothesis. Testing multiple templates allows us to evaluate the stability of the recovered Galactic signal under different assumptions of CR propagation.

The Galactic signal is modeled using four emission templates: the \textit{Fermi-LAT $\pi^0$}, KRA$\gamma^5$, KRA$\gamma^{50}$, and CRINGE templates. The \textit{Fermi}-LAT $\pi^{0}$ template~\autocite{Fermi-LAT:2012edv} is based on diffuse $\gamma$-ray observations and predicts a relatively broad emission profile along the GP. The KRA$_\gamma^5$ and KRA$_\gamma^{50}$ templates~\autocite{Gaggero:2015xza} include spatially dependent cosmic-ray transport, producing harder spectra in the inner Galaxy and emission more concentrated toward the Galactic Centre. The CRINGE template~\autocite{Schwefer:2022zly} is derived from a global fit to cosmic-ray data and includes an unresolved Galactic-source component, giving predictions that lie between the \textit{Fermi}-LAT $\pi^{0}$ and KRA$_\gamma$ cases. Further details of the individual templates are provided in the Supplementary Information.

The likelihood is parametrized in terms of the expected number of signal events, $n_s$, which is related to the overall flux normalization and acts as the sole fit parameter~\autocite{Braun:2009wp}. Signal PDFs are constructed from MC, assuming a model flux with selection cuts applied. 

As a simplification, the energy and directional PDFs that enter the likelihood are factorized~\autocite{IceCube:2023ame} by assuming a sky-averaged spectrum for the KRA$_\gamma$ and CRINGE templates. Background PDFs are modelled using a data-driven approach that exploits symmetries in the detector; 
any signal contamination in data is accounted for by a signal-subtraction term~\autocite{IceCube:2015rnn}. Similarly, the TS distribution under the null hypothesis, which is used to evaluate statistical significance, is obtained by repeatedly randomising the right ascension of all data events and evaluating the TS for each iteration. Additional details are provided in the Methods section.

\begin{figure*}[ht!]
    \centering
        \includegraphics[width=0.95\textwidth]{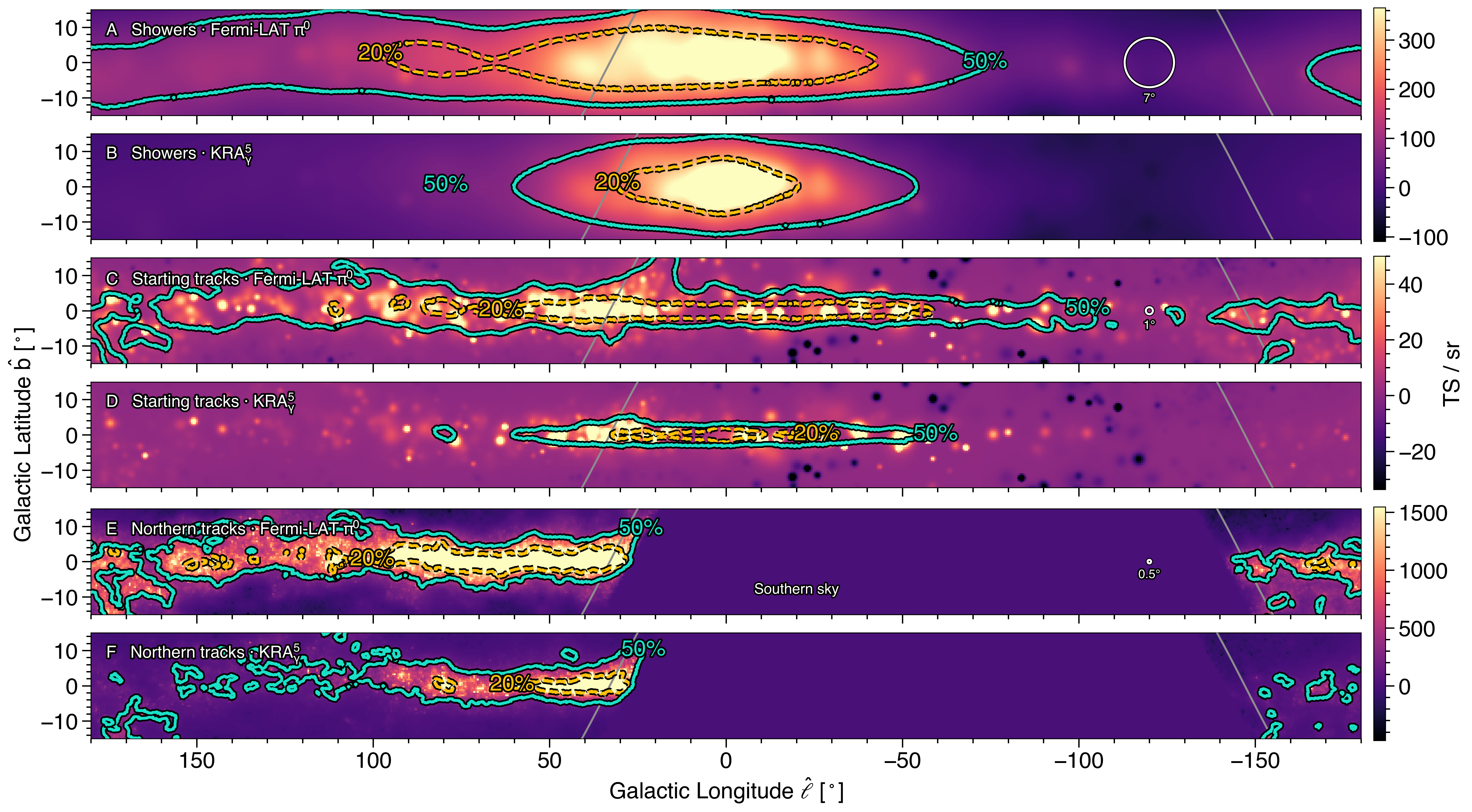}
    \caption{\textbf{Test statistic (TS) distributions in Galactic coordinates.} Colour maps show the TS per steradian for the shower (A, B), starting-track (C, D) and northern-track (E, F) samples, as indicated by the colour bar (unique to each sample). Contour lines indicate \SI{20}{\%} and \SI{50}{\%} containment regions of the signal probability density function (PDF) for the corresponding template, assuming a typical angular uncertainty for an event from the sample (white circle, \ang{7} for showers, \ang{1} for starting tracks and \ang{0.5} for through-going tracks). Grey lines mark $\delta = \ang{0}$ in equatorial coordinates. The TS distributions depend on the assumed Galactic template: the \textit{Fermi}-LAT $\pi^0$ model (A, C, E) predicts
    a broader profile along Galactic longitude, while KRA$_\gamma^5$ (B, D, F) peaks more prominently
    toward the Galactic Centre. Finer features are visible for the track samples due to their superior angular resolution. The lack of sensitivity from through-going tracks in the northern sky, within the Galactic longitude range from $\ang{-150}$ to $\ang{20}$, is also apparent in the bottom panels. Maps for the CRINGE  
    and KRA$_\gamma^{50}$ results are shown in Extended Data Fig.~\ref{fig:ts_stack_method}.}
    \label{fig:iceman_TS_map}
\end{figure*}

In 12 years of data, across the four models tested, the \textit{Fermi}-LAT $\pi^0$ template yields the maximum TS, which is used to calculate the global $p$-value. The significance is driven by the shower sample, with a local significance of \SI{5.29}{\textsigma}. Compared to the previous results~\autocite{IceCube:2023ame}, the difference is attributable to improved ice modelling in simulation~\autocite{tc-18-75-2024,IceCube:2023qua} and reconstruction~\autocite{IceCube:2024csv}, and the increase in livetime. The track samples show less-significant excesses, indicating only a hint of diffuse Galactic emission. Their best-fit normalizations are consistent within uncertainties to that of the shower sample. Together, the local significance increases to \SI{5.97}{\textsigma} in the combined analysis. Model consistency checks are shown in Extended Data Fig.~\ref{fig:roundtrip}. The trend in significance in the data agrees better with pseudo-experiments generated assuming the \textit{Fermi}-LAT $\pi^0$ and CRINGE templates; injections based on the KRA$_\gamma$ templates tend to produce a different ordering of significances. A detailed breakdown of the fitted normalizations and significances for the individual samples is provided in Supplement, Tab.~1. Compared with the TS distribution under the null hypothesis, and accounting for the look-elsewhere effect, the post-trial one-sided Gaussian-equivalent significance of the observed signal is \SI{5.7}{\textsigma}.

The fitted normalizations differ substantially between the tested models. The \textit{Fermi}-LAT $\pi^0$ template requires a normalization factor of $\asym{4.99}{0.90}{0.86}$, indicating that the nominal model underestimates the observed neutrino flux. In contrast, the KRA$\gamma^{5}$ and KRA$\gamma^{50}$ templates prefer normalizations of $\asym{0.53}{0.14}{0.12}$ and $\asym{0.37}{0.10}{0.09}$, respectively, corresponding to nominal predictions above the flux level preferred by the data. The CRINGE template prefers a normalization factor of $\asym{0.91}{0.18}{0.18}$, consistent with its nominal prediction within uncertainties. These varying scale factors reflect the combined effect of adjusting the assumed gas distribution, Galactic CR density and spectrum, as well as unresolved Galactic neutrino sources in order to describe the neutrino data. 
While the CRINGE result indicates agreement with nominal predictions in the energy range probed here, it does not identify which of these ingredients drives the agreement.

The colour maps in Fig.~\ref{fig:iceman_TS_map} show the TS density (the total TS contribution within each HEALPix~\autocite{Gorski:2004by} pixel, normalized by its solid angle) centred on the GP for two of the tested templates. Contributions from the shower (panels A, B), starting-track (C, D), and northern-track (E, F) samples are shown, alternating between the \textit{Fermi}-LAT $\pi^0$ and KRA$\gamma^{5}$ templates. Additionally, contour lines indicate \SI{20}{\%} and \SI{50}{\%} containment regions of the signal PDF for the corresponding template, assuming a typical angular uncertainty for an event from the sample (white circles). Finer features are visible in the lower two rows due to the superior angular resolution of tracks compared to showers. 
Extended Data Fig.~\ref{fig:ts_stack_method} shows maps for the CRINGE and KRA$_\gamma^{50}$ templates. Since the same dataset is fitted in all cases, differences between the TS distributions should not be interpreted as independent measurements, but rather as interpretations of the same data assuming different Galactic-emission templates derived from alternative cosmic-ray propagation models.

\begin{figure*}[ht!]
\centering
\includegraphics[width=0.9\textwidth]{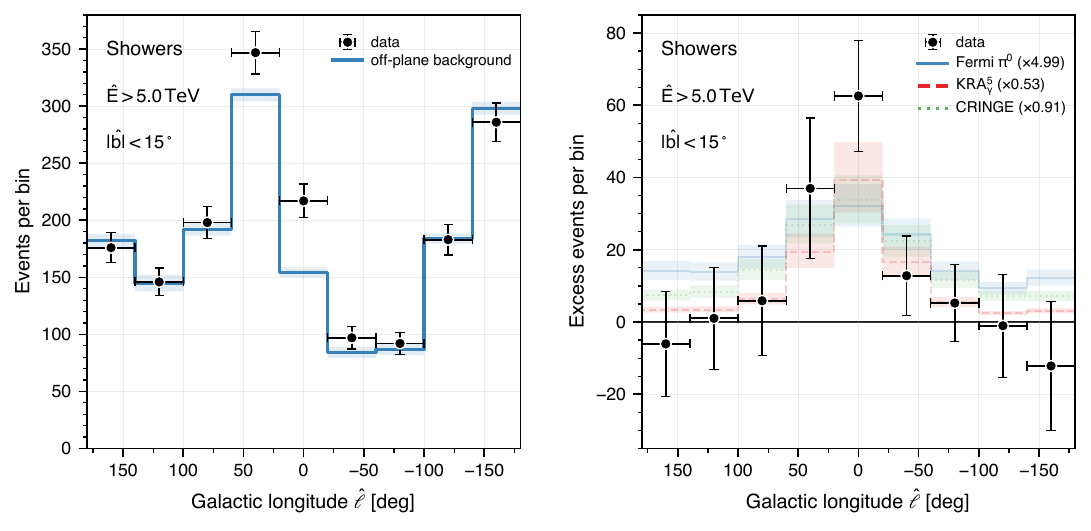}
\caption{
\textbf{Distribution and background-subtracted residuals of shower events along Galactic longitude.}
The left panel shows observed counts (black points) for shower events with reconstructed energy $\hat{E}>\SI{5}{\tera \eV}$ and Galactic latitude $|\hat{b}|<15^\circ$, with error bars corresponding to the uncertainty from Poisson statistics.
The blue histogram shows data-derived background expectations, estimated from off-plane regions along the same declination; the shaded band corresponds to its statistical uncertainty.
The right panel shows residuals (black points) after background subtraction, $N_{\mathrm{obs}}(\hat{\ell})-N_{\mathrm{bkg}}(\hat{\ell})$, with uncertainties obtained via error propagation. Coloured lines show signal expectations from three of the tested templates, scaled to their best-fit normalizations from the likelihood analysis; shaded bands indicate the corresponding \SI{1}{\textsigma} uncertainty on the normalization. An excess is visible, concentrated toward the inner Galaxy, that is broadly consistent in scale and shape with diffuse Galactic emission models. Model expectations are overlaid for visual comparison only; no fit is performed to the binned points shown here.
}
\label{fig:shower-longitude-excess}
\end{figure*}

While the TS maps highlight regions where each template is preferred over the null hypothesis, they do not show how well a given model reproduces the data. To corroborate and visualize the signal excess, we define on-plane and off-plane regions using a Galactic-latitude bound, $b_{\max}$, and an energy threshold, $E_{\min}$. Events with $|\hat{b}| > b_{\max}$ and $\hat{E} > E_{\min}$ are treated as background (off-plane), while events with $|\hat{b}| < b_{\max}$ and $\hat{E} > E_{\min}$ contain both signal and background (on-plane). A minimum energy threshold helps to suppress atmospheric backgrounds, which dominate at lower energies, and both threshold values are obtained by MC-based optimizations as detailed in the Methods section. The selections are optimized separately for showers and tracks using MC pseudo-experiments, giving $b_{\max}=15^\circ$, $E_{\min}=\SI{5}{\tera\eV}$ for showers and $b_{\max}=8.5^\circ$, $E_{\min}=\SI{0.1}{\tera \eV}$ for tracks. The background subtraction is data-driven and independent of the signal model. To ensure sufficient statistics while remaining agnostic to small-scale structure, 
Galactic longitude $\hat{\ell}$ was binned coarsely in intervals of \ang{40}.

The left panel of Fig.~\ref{fig:shower-longitude-excess} shows the observed counts (black points) and data-derived background estimates (blue lines) for shower events along Galactic longitude. Error bars correspond to uncertainties from Poisson statistics on the observed counts, and the shaded blue band indicates the statistical uncertainty on the estimated background, derived from data and scaled to the on-plane solid angle within each bin. The right panel shows residuals (black points) after background subtraction, $N_{\rm obs}(\hat{\ell})-N_{\rm bkg}(\hat{\ell})$, with uncertainties obtained via error propagation. Coloured lines correspond to signal expectations from three of the tested templates, scaled to their best-fit normalizations from the likelihood analysis; shaded bands indicate the corresponding \SI{1}{\textsigma} uncertainty on the measured normalization. These model expectations are provided as a visual aid only, and it is important to note that no fit is performed to the data points. An excess is visible in the data, concentrated toward the inner Galaxy, that is broadly consistent in scale and shape with the models. Distributions and residuals for track events, obtained under the same construction, are shown in Extended Data Fig.~\ref{fig:track-longitude-residuals}. A weaker excess in the track data is visible in the northern sky.

As seen in Fig.~\ref{fig:shower-longitude-excess}, the largest excess is found in the central longitude bin, $|\hat{\ell}|<20^\circ$. 
In this bin, we observe $\num{217}$ shower events, compared with an estimated background of $\num{154.4 \pm 4.1}$, resulting in a residual of $\num{62.6 \pm 15.3}$ events.
The Galactic templates, with best-fit normalizations applied, predict $30$--$40$ excess events in the same bin. While the measured excess is higher than these expectations, the difference is not statistically significant within the uncertainties shown.

\begin{figure}[ht!]
    \centering
    \includegraphics[width=0.93\linewidth]{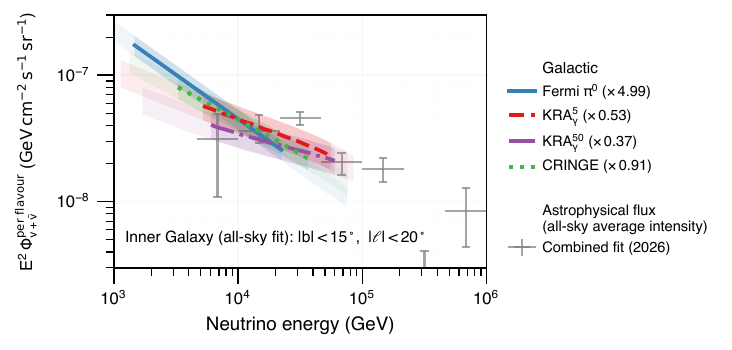}   
    \caption{\textbf{Diffuse Galactic neutrino spectra averaged over the Inner Galaxy.} We present the best-fit per-flavour differential flux for four Galactic emission models: \textit{Fermi}-LAT~$\pi^0$ (blue), KRA$_\gamma^{5}$ (red), KRA$_\gamma^{50}$ (purple), and CRINGE (green). The model predictions are scaled by the best-fit normalization obtained from the global likelihood analysis. The spectra are shown for the inner Galactic region \((|b|\le15^\circ,\, |\ell|\le20^\circ)\).
    The sensitive energy range is denoted by the shaded bands; horizontal spans indicate 68\% (90\%) regions in darker (lighter) shades. Vertical spans on the spectra denote 68\% uncertainties from a unified Feldman–Cousins construction~\autocite{Feldman:1997qc}, which includes detector systematics. See Methods for details. 
    The Galactic neutrino spectra are compared to IceCube measurements of the all-sky-averaged astrophysical neutrino flux using 
    a combination of tracks and contained showers~\autocite{IceCube:2025tgp} (Combined Fit), derived under the assumption of isotropy over the full sky.
}
 \label{fig:flux_plots}
\end{figure}

Assuming the templates described earlier, we next characterize the associated neutrino spectra and compare them to measurements of the isotropic astrophysical neutrino flux. As the likelihood is dominated by the inner Galaxy, the average flux over the entire sky is not representative of the region that drives the fit. Instead, here we focus on the template-predicted flux in the inner region, scaled by the best-fit normalization parameter from the all-sky measurement. Figure~\ref{fig:flux_plots} shows the spectra for the four Galactic emission templates, where the scaled, template-predicted flux within the inner Galaxy is shown, averaged over that region. The inner Galaxy is defined analogous to the shower-optimized region in Fig.~\ref{fig:shower-longitude-excess}. The convergence of the four post-fit spectra within the inner-Galaxy region, despite their substantially different pre-fit predictions (Extended Data Fig.~\ref{fig:model_fits}), reflects the fact that current data constrain the overall emission level there but cannot yet distinguish between the competing spatial and spectral assumptions of the tested CR transport models.

For comparison, Fig.~\ref{fig:flux_plots} also shows the inferred Galactic intensity alongside the isotropic astrophysical neutrino flux measured by IceCube using a combination of tracks and contained showers~\autocite{IceCube:2025tgp} (per-solid-angle rather than integrated flux over sky). The fitted flux from the inner-Galaxy region reaches a level comparable to the all-sky averaged astrophysical intensity. Note that this region spans \(\simeq\SI{0.36}{\steradian}\), or about \(\SI{3}{\percent}\) of the sky, and therefore does not imply that the IceCube diffuse astrophysical flux measurement is dominated by Galactic emission. Integrated over the full sky, the fitted Galactic component corresponds to approximately \(\SI{10}{\percent}\) of the total astrophysical flux at \(E_\nu\sim\SI{20}{\tera\eV}\), used here as a reference energy for comparison since it lies in the sensitive energy range of this analysis, and where the all-sky astrophysical flux is well constrained by recent IceCube measurements~\autocite{IceCube:2024fxo,IceCube:2025tgp}. Similarly, differences in spectral shape, in particular at lower energies, is not evidence of a tension between Galactic and the all-sky fluxes at this time. Future work can provide a more detailed decomposition of the Galactic energy spectra and increased precision on the all-sky flux in this energy regime.

Taken together, the measurements presented here, obtained using a 12-year, all-flavour IceCube dataset, establish a spatially non-uniform Galactic contribution to the high-energy neutrino sky. The signal appears as an extended structure aligned with the GP and concentrated in the inner Galaxy, where an excess of greater than 60 events is observed. The present data do not yet uniquely separate diffuse interstellar emission from unresolved source populations or other subdominant contributions, for instance, potential dark-matter-induced neutrino emission. Identifying and constraining these components will require improved angular resolution, higher statistics, and joint comparisons with high-energy $\gamma$-ray observations. In particular, future measurements will help distinguish diffuse interstellar emission from unresolved Galactic source populations~\autocite{Ambrosone:2023hsz,Sun:2023ibg,Gagliardini:2024een,Fang:2024fyd,Peretti:2024ecg,Carpio:2025arz,Moghadam:2026jmy,Menchiari:2026ney}. Looking further ahead, precise measurements of the Galactic neutrino spectrum and flavour composition, together with observations from northern-hemisphere neutrino observatories such as KM3NeT~\autocite{KM3NeT:2024paj}, may also enable tests of neutrino properties over kiloparsec baselines, complementing terrestrial and extragalactic measurements~\autocite{Beacom:2002vi,Kobayashi:2000md,Ioka:2014kca,Ng:2014pca}.

\printbibliography[notkeyword=methods, title={References}]

\section*{Methods}
\input{method}
\printbibliography[keyword=methods, title={Methods references}]




\subsection*{IceCube Collaboration$^{\ast}$:}
\input{authorlist}

\section*{Funding}
The authors gratefully acknowledge the support from the following agencies and institutions:
USA {\textendash} U.S. National Science Foundation-Office of Polar Programs,
U.S. National Science Foundation-Physics Division,
U.S. National Science Foundation-EPSCoR,
U.S. National Science Foundation-Office of Advanced Cyberinfrastructure,
Wisconsin Alumni Research Foundation,
Center for High Throughput Computing (CHTC) at the University of Wisconsin{\textendash}Madison,
Open Science Grid (OSG),
Partnership to Advance Throughput Computing (PATh),
Advanced Cyberinfrastructure Coordination Ecosystem: Services {\&} Support (ACCESS),
Frontera and Ranch computing project at the Texas Advanced Computing Center,
U.S. Department of Energy-National Energy Research Scientific Computing Center,
Particle astrophysics research computing center at the University of Maryland,
Michigan State University,
Astroparticle physics computational facility at Marquette University,
NVIDIA Corporation,
and Google Cloud Platform;
Belgium {\textendash} Funds for Scientific Research (FRS-FNRS and FWO),
FWO Odysseus and Big Science programmes,
and Belgian Federal Science Policy Office (Belspo);
Germany {\textendash} Bundesministerium f{\"u}r Forschung, Technologie und Raumfahrt (BMFTR),
Deutsche Forschungsgemeinschaft (DFG),
Helmholtz Alliance for Astroparticle Physics (HAP),
Initiative and Networking Fund of the Helmholtz Association,
Deutsches Elektronen Synchrotron (DESY),
and High Performance Computing cluster of the RWTH Aachen;
Sweden {\textendash} Swedish Research Council,
Swedish Polar Research Secretariat,
National Academic Infrastructure for Supercomputing in Sweden (NAISS), and Knut and Alice Wallenberg Foundation; 
European Union {\textendash} EGI Advanced Computing for research;
Australia {\textendash} Australian Research Council;
Canada {\textendash} Natural Sciences and Engineering Research Council of Canada,
Calcul Qu{\'e}bec, Compute Ontario, Canada Foundation for Innovation, WestGrid, and Digital Research Alliance of Canada;
Denmark {\textendash} Villum Fonden, Carlsberg Foundation, and European Commission;
New Zealand {\textendash} Marsden Fund;
Japan {\textendash} Japan Society for Promotion of Science (JSPS), Ministry of Education, Culture, Sports, Science and Technology (MEXT), and Institute for Global Prominent Research (IGPR) of Chiba University;
Korea {\textendash} National Research Foundation of Korea (NRF);
Switzerland {\textendash} Swiss National Science Foundation (SNSF).\\

\section*{Author contributions}
The IceCube Collaboration designed, constructed and now operates the IceCube Neutrino Observatory. Data processing and calibration, Monte Carlo simulations of the detector and of theoretical models, and data analyses were performed by a large number of collaboration members, who also discussed and approved the scientific results presented here. The IceCube collaboration acknowledges the substantial contributions to this manuscript from the University of Wisconsin--Madison and Drexel University. The manuscript was reviewed by the entire collaboration before publication, and all authors approved the final version.\\



\clearpage
\section*{Extended data figures}
\input{ed}

\clearpage
\input{supplementary}

\end{document}

%% file: method.tex
\subsection*{Description of datasets}
\label{sec:methods_dataset}

Event morphologies in IceCube correlate with neutrino-nucleon interactions and the subsequent processes of charged, secondary particles. In $\nu_\mu$ charged-current (CC) interactions, a muon is produced that, at energies above $\sim\SI{1}{\tera \eV}$,  can travel km-scale distances through ice. Such event signatures in IceCube are referred to as ``tracks''. In contrast, $\nu_e$ CC interactions typically produce more compact electromagnetic and hadronic particle showers with a characteristic longitudinal length of $\mathcal{O}(10\,\mathrm{m})$. In this paper, we use ``shower'' when referring to both the underlying physical process of particle showers and such event morphologies. The terminology ``cascades'' has also been used in the literature to refer to the latter. In $\nu_\tau$ CC interactions, depending on the $\tau$ energy and decay products, the event signature can appear either as two showers separated by some distance (for longer decay lengths and leptonic decays into electrons or hadronic decays), as a track (for leptonic decays into muons), or virtually indistinguishable from $\nu_e$ CC. Finally, in neutral-current interactions only a fraction of the neutrino energy remains visible in the nucleon fragments, leaving a shower signature.

Selections optimized for astrophysical neutrinos have separately targeted specific event morphologies. The relevant subsamples used in this analysis are described below and include data collected since 2010, when IceCube was still in its final construction season before completion of the full 86-string detector configuration in 2011. The through-going track sample~\autocite{Abbasi:2021qfz,IceCube:2022der} (NT) contains data collected between June 2010 and November 2023, while the shower sample~\autocite{IceCube:2023ame} (DNNC) and the enhanced starting track sample~\autocite{IceCube:2025zyb} (ESTES) contain data collected between May 2011 and November 2023.
Standard data-quality requirements are applied throughout.

The DNNC sample targets neutrino-induced particle showers with interaction vertices contained within the instrumented volume of IceCube or within 300~m of its boundary. The dominant backgrounds are atmospheric muons and atmospheric muon neutrinos, both of which predominantly produce track-like topologies. Event classification is performed using a multistage machine-learning approach that combines deep neural networks (DNNs) and boosted decision trees (BDTs). In the early stages of the selection, fast DNN classifiers are applied using low-level, per-optical-module summary statistics, including the total collected charge, the time of the first detected pulse, and the charge-weighted temporal spread of pulses. These classifiers efficiently suppress the high-rate of atmospheric muon background. At later stages, more computationally intensive neural networks are applied to further refine the separation between shower-like and track-like events using detailed spatio-temporal information. In the final stage of the selection, BDTs combine the outputs of the DNN classifiers with a set of high-level topology variables to reject residual track-like backgrounds and define the final shower sample used in the analysis, resulting in a neutrino purity of greater than 90\%.

The NT sample uses distinct selection strategies for through-going and starting events. For upward-going tracks, the Earth acts as an effective shield against atmospheric muons, which are stopped prior to reaching the detector. Restricting the selection to tracks originating from the northern sky therefore yields a high-statistics sample with neutrino purity above 99.8\%, corresponding to very low atmospheric-muon contamination. The selected events are nevertheless dominated by atmospheric neutrinos, with astrophysical neutrinos contributing primarily at high energies.
Additional quality criteria are applied to retain well-reconstructed track events, and a BDT is used to suppress cascade-like topologies, which have less reliable angular reconstruction. Finally, the energy reconstruction is improved by a neural network~\autocite{IceCube:2022der} below \SI{1}{\tera \eV}, an energy regime where muons become minimum ionizing.

The ESTES sample targets all-sky neutrino-induced muon tracks with interaction vertices inside the detector volume. To distinguish these events from the much higher rate of atmospheric muons, a dynamic veto strategy is employed instead of a fixed geometric veto. For each event, the reconstructed track direction and time of the earliest detected photons are used to define an event-specific veto region upstream of the interaction point under the through-going muon hypothesis. The observed hit pattern in this region is evaluated for consistency with a muon that enters the detector, and events compatible with such a hypothesis are rejected. A BDT using the dynamic-veto information and additional high-level topology variables is then included as part of the final selection to reject remaining background-like events. The neutrino purity is about 92\% (99\%) in the southern (northern) sky~\autocite{Mancina:2021jbk}. 

Through-going tracks typically provide the best angular resolution, but the through-going track sample is less effective in the southern sky because of atmospheric-muon backgrounds. This makes showers and starting tracks especially important for diffuse Galactic neutrino emission, which is expected to peak near the GC~\autocite{Fermi-LAT:2012edv,Gaggero:2015xza,Schwefer:2022zly} at a declination of $\delta \simeq -29^\circ$~\cite{Reid:2004rd}. These event morphologies provide stronger southern-sky muon rejection.

Since the datasets are not mutually exclusive, events that were selected by multiple samples, or overlapping events, are flagged and removed such that they appear in only one subsample in the combined dataset. The decision strategy was based on Monte Carlo (MC) studies of their angular and energy resolutions, which can differ between samples due to the different reconstruction algorithms used. Duplicate events in overlapping samples are removed as follows: events in ESTES and DNNC are removed from ESTES; events in DNNC and NT are removed from DNNC; events in NT and ESTES are removed from NT. Events in all three datasets are kept in NT. For ESTES, the removed events constitute $\sim2.4\%$ of the original. For DNNC and NT the contribution is less than 1\%.

Extended Data Fig.~\ref{fig:data_distributions} shows the event counts in data for all three selections. Through-going tracks dominate the total event statistics because the effective detection volume for tracks is much larger than the instrumented detector volume. The through-going-track sample is dominated in total event count by atmospheric neutrinos. The astrophysical component contributes at the sub-percent level at TeV energies and becomes comparable to the atmospheric-neutrino component only at reconstructed energies of $\mathcal{O}(100~\mathrm{TeV})$.

In the southern sky, atmospheric neutrinos produced in cosmic-ray air showers are often accompanied by energetic muons from the same parent interactions. Because both the shower and starting-track selections reject muons entering the detector, atmospheric neutrinos from the same air shower are often also removed, a suppression known as the atmospheric ``self-veto'' effect~\autocite{Gaisser:2014bja,Arguelles:2018awr}.

\subsection*{Reconstruction improvements and angular uncertainty estimation for showers}

The inclusion of several updates in the description of the South Pole ice leads to significantly improved agreement for IceCube calibration data. Ice birefringence~\autocite{tc-18-75-2024}, ice layer undulations~\autocite{IceCube:2023qua}, and a recalibration of how the refrozen ice within the drill hole modifies the angular sensitivity of IceCube sensors~\autocite{IceCube:2013llx} are now incorporated in the modelling of in-ice particle showers for event reconstruction~\autocite{IceCube:2024csv}. With birefringence, a more uniform azimuthal distribution of shower-like events in 12 years of IceCube data is obtained compared to that from the previous Galactic plane (GP) analysis~\autocite{IceCube:2023ame}. Here, the azimuthal angle is taken from a detector-centred, spherical coordinate system and thus is expected to be largely uniform. A more accurate description of ice layer undulations is especially important for shower-like events produced outside, or near the edge of, the instrumented volume, while the angular sensitivity modification can affect the zenith (polar) angle reconstruction of particle showers. 

The red, dashed contour in Extended Data Fig.~\ref{fig:iceman_azimuth} shows the azimuthal distribution using the prior hybrid reconstruction, which relied on a DNN to predict expected Cherenkov light yields at each sensor in order to perform maximum likelihood fits. Event rate excesses (deficits) in data are visible along (perpendicular to) the direction of the glacial ice flow with respect to the detector, an artefact that was a result of training the DNN on a now outdated ice model without birefringence effects, or the updates in ice layer undulations and hole-ice angular sensitivity. The blue, solid contour shows the azimuthal distribution for shower-like events used in this analysis.

The updated simulation also yields improved directional resolution across the full sensitive energy range, as shown in the right panel of Extended Data Fig.~\ref{fig:iceman_reco_sim}. Evaluating both reconstructions on the latest MC further reveals a systematic zenith bias in the reconstruction used in the prior analysis~\autocite{IceCube:2023ame} that was not apparent in simulation used at the time. The zenith angle, $\theta$, is defined as the polar angle in a detector-centric spherical coordinate system, with $\theta$ less (greater) than $\ang{90}$ corresponding to arrival directions from above (below) the horizon.
As shown in left panel of Extended Data Fig.~\ref{fig:iceman_reco_sim}, the prior reconstructed zenith is pulled towards the horizon; events are preferentially reconstructed toward $\theta \approx 90^\circ$, with the median residual $\Delta\theta = \hat{\theta} - \theta_{\mathrm{true}}$ reaching up to $\sim 6^\circ$ in the southern hemisphere and $\sim 10^\circ$ in the northern hemisphere. The updated reconstruction exhibits less bias across the entire zenith range, where the small deviations at $\theta_{\mathrm{true}} = \ang{0}$ and $\ang{180}$ are expected as an artefact of the choice of coordinate system.

To predict the per-event angular uncertainty, we utilize the Light Gradient Boosting Machine (LightGBM) framework~\autocite{NIPS2017_6449f44a} to train a Gradient Boosted Decision Tree (GBDT). The GBDT was trained on MC using a variety of features, including the reconstructed position, direction and energy, as well as goodness-of-fit metrics~\autocite{IceCube:2025ixg}. The loss function is the negative logarithm of 
a von Mises--Fisher (vMF) distribution on the sphere (up to a normalization term), with concentration parameter $\kappa$ representing the angular uncertainty $\sigma\equiv 1/\sqrt{\kappa}$, which is the target variable. Note that for large $\kappa$, $\sigma$ can be interpreted as the scale parameter of a 2D Gaussian.
In practice, a more numerically stable parametrisation in terms of $\eta=\ln(\kappa)$ is used that avoids negative $\kappa$,
\begin{equation}\label{eqn:loss_function}
    L(\psi;\eta) = e^{\eta} +\ln(1-e^{-2e^{\eta}})-\eta-e^{\eta}\cos(\psi),
\end{equation}
where $\psi$ is the angle between reconstructed and true directions. The gradient of the loss function is also used in the GBDT training and was calculated to be
\begin{equation}
    \nabla_\eta(L(\psi;\eta)) = e^{\eta}\coth\left(e^{\eta}\right)-1-e^{\eta}\cos(\psi).
\end{equation}
The predicted $\hat{\sigma}$ for each event then enters the signal probability density function (PDF) used in the likelihood evaluation, as described in the next section.

Finally, the best-fit reconstructed energy of shower-like events now reflects the latest calibrations of photon-detection efficiencies. The overall impact is a slightly higher reconstructed energy over the dataset, on the order of \SIrange{5}{10}{\%}. Uncertainties on this and other detector-related systematics are taken into account in the flux measurement, as described later.

\subsection*{Likelihood ratio construction}

One distinction of this analysis relative to previous IceCube Galactic-plane neutrino searches~\autocite{IceCube:2023ame} is the inclusion of multiple event samples within a single likelihood framework. Similar combined-sample approaches have been developed in diffuse neutrino analyses targeting the extragalactic flux~\autocite{IceCube:2015gsk,IceCube:2025tgp,IceCube:2025gna}, which typically rely on detailed forward modelling of atmospheric neutrino and muon backgrounds using MC.
Similar strategies for combining multiple event classes have been used in point-source searches~\autocite{IceCube:2025lev}, though the likelihood implementation in this work is adapted to the case of extended source emission.

In this analysis, the background PDF construction is data-driven instead of MC-driven. Taking advantage of IceCube's location at the South Pole, where Earth's rotation changes only changes only the local azimuthal angle of a celestial object while leaving its zenith angle unchanged, the test-statistic (TS) distribution under the null hypothesis is obtained by randomizing the right ascensions, $\alpha$, of all data events. The analysis employs a joint likelihood over multiple event samples. Here, we describe the construction of the signal and background PDFs, define the TS based on the likelihood ratio, and detail how the fit parameter is related to the overall model normalization.

For each of the four tested models of our Galaxy, the total number of signal events enters the likelihood as the parameter of interest; the shape of the PDF in direction and energy is not fitted. The likelihood is constructed from a joint probability over all events in each of the datasets described above, where the probability of each event is modelled as a two-component mixture of signal and background probabilities~\autocite{Braun:2009wp}.

For each dataset $j$, the expected number of signal events under a given flux model is
\begin{equation}
\lambda_s^j
=
\int \mathrm{d}\bOmega\,\mathrm{d}E\,\mathrm{d}t\;
\Phi(\bOmega,E)\,
A_{\mathrm{eff}}^j(\delta,E),
\label{eq:lambdas}
\end{equation}
where $\Phi$ is the predicted flux and $A^j_\mathrm{eff}$ is an efficiency term corresponding to the cross-sectional area of a hypothetical detector that is \SI{100}{\%} efficient at detecting the flux. Both terms have a dependence on the neutrino direction, $\bOmega$ (in equatorial coordinates $\delta$ is the declination), and energy, $E$, and the integration over those as well as the livetime, $t$, results in the number of expected events in the dataset, $\lambda_s^j$.  An equal flavour ratio at Earth of $\nu_e:\nu_\mu:\nu_\tau = 1:1:1$ is assumed, as expected from neutrino oscillations over Galactic distances. Changing the flavour ratio would change the number of expected signal events in subsample $j$, $\lambda_s^j$, for a given flux. The total number of expected events is just the sum over all datasets, $\lambda_s = \sum_j{\lambda_s^j}$. In practice, such integrals are challenging to perform analytically and are instead calculated based on MC simulations.
To simplify the calculation, and following earlier procedures~\autocite{IceCube:2023ame}, the signal PDF is factorized into separate directional and energy PDFs, which makes the eventual likelihood computation feasible. Alternative approaches that retain the full joint spatial--energy dependence have also been explored~\autocite{IceCube:2023hou}. For each event, $i$, in dataset, $j$, the signal PDF is given by
\begin{equation}
\label{eq:sxe}
\mathcal{S}^j(\hat{\bOmega}_i,\hat{E}_i; \hat{\sigma}_i)
=
\mathcal{N}^j(\hat{\bOmega}_i; \hat{\sigma}_i)\,
\mathcal{E}^j(\hat{E}_i| \hat{\delta}_i),
\end{equation}
where the hat notation here and throughout the paper indicates that the PDFs are given in terms of reconstructed quantities, $\hat{\sigma}_i$ is the event-dependent angular uncertainty, and $\hat{\delta}_i$ corresponds to the declination, on which the energy PDF, $\mathcal{E}^j$, is conditioned in order to account for the declination-dependent detector response. The directional PDF, \(\mathcal{N}^j\), is first constructed from $\Phi$ and $A_\mathrm{eff}^j$, then marginalized over the energy to get
\begin{equation}
\mathcal{N}^j(\bOmega)
=
\frac{
\int \mathrm{d}E\,\mathrm{d}t\;
\Phi(\bOmega,E)\,
A_{\mathrm{eff}}^j(\delta,E)}
{\lambda_s^j}.
\label{eq:spdf}
\end{equation}
Next, the event-dependent angular uncertainty is accounted for by convolving \(\mathcal{N}^j\) with a vMF kernel (the spherical equivalent of a Gaussian kernel) parametrized by the predicted uncertainty of the event, $\hat{\sigma}_i$, resulting in a directional PDF, $\mathcal{N}^j(\hat{\bOmega}_i; \hat{\sigma}_i)$, that depends on $\hat{\sigma}_i$.

The energy PDF, \(\mathcal{E}^j\), is derived from the solid-angle-integrated spectrum,
\begin{equation}
\Phi_{\mathrm{avg}}(E)
= 
\int \Phi(\bOmega,E)\,\mathrm{d}\bOmega
\end{equation}
and relies on MC to account for differences between true and reconstructed quantities. The MC is, in essence, a stochastic ensemble of points sampled from, and hence approximating, a high-dimensional PDF. It allows for evaluation of the conditional probability $\rho_E^j(\hat{E}|E, \delta, \bm{\Theta})$ to obtain reconstructed energy $\hat{E}$ given true properties $E$, $\delta$, and $\bm{\Theta}$, where $\bm{\Theta}$ represents other relevant event properties such as the neutrino flavour and interaction channel. As each dataset uses its own reconstruction routine, $\rho_E^j$ is dataset-dependent as indicated by the superscript $j$. Similarly, let $\rho_\delta^j(\hat{\delta}|E, \delta, \bm{\Theta})$ represent the conditional probability of obtaining reconstructed declination $\hat{\delta}$. Then, the energy PDF can be expressed as
\begin{equation}
\mathcal{E}^j(\hat{E}_i| \hat{\delta}_i)
= 
\frac{
\int \Phi_{\mathrm{avg}}(E)A_{\mathrm{eff}}^j(\delta, E)\rho_E^j(\hat{E}_i|E, \delta, \bm{\Theta})\rho_\delta^j(\hat{\delta}_i|E, \delta, \bm{\Theta}) \,\mathrm{d}E\,\mathrm{d}\bOmega\,\mathrm{d}\bm{\Theta}
}
{
\int \Phi_{\mathrm{avg}}(E)A_{\mathrm{eff}}^j(\delta, E) \rho_\delta^j(\hat{\delta}_i|E, \delta, \bm{\Theta}) \,\mathrm{d}E\,\mathrm{d}\bOmega\,\mathrm{d}\bm{\Theta}
}.
\label{eq:epdf}
\end{equation}

The background PDF construction is data driven. Treating the observed data as some combination of signal and background, and assuming the background is independent of right ascension (RA), the observed data PDF can be expressed as
\begin{equation}\label{eq:sigsub}
    \bar{\mathcal{D}}(\hat{\delta}_i, \hat{E}_i; \hat{\sigma}_i) = f \bar{\mathcal{S}}(\hat{\delta}_i, \hat{E}_i; \hat{\sigma}_i) + (1-f) \mathcal{B}(\hat{\delta}_i, \hat{E}_i),
\end{equation}
where $f$ is the fraction of signal events in the dataset, $\mathcal{B}$ the underlying background PDF, and the overline indicates that these are marginalized over RA~\autocite{IceCube:2015rnn}. Following earlier constructions~\autocite{Braun:2009wp}, the likelihood of observing $n_s$ signal events across all datasets is
\begin{equation}\label{eq:llh}
\mathcal{L}(n_s) = \prod_{j} \left[ \prod_{i}^{N^j} 
\frac{r^j n_s}{N^j} \left( 
\mathcal{S}^j(\hat{\bOmega}_i, \hat{E}_i; \hat{\sigma}_i) - \bar{\mathcal{S}}^j (\hat{\delta}_i, \hat{E}_i; \hat{\sigma}_i) \right)+ 
\bar{\mathcal{D}}^j(\hat{\delta}_i, \hat{E}_i; \hat{\sigma}_i) \right],
\end{equation}
where $r^j \equiv \lambda_s^j / \lambda_s$ and $N^j$ the total number of events in dataset $j$. Note that the total number of signal events, $n_s$, is the sole parameter, and the relative contribution of each dataset is determined by $r^j$. The test statistic (TS) is defined as the logarithm of the ratio between the maximum likelihood of the alternative hypothesis to that of the null hypothesis,
\begin{equation}\label{eq:TS}
\mathrm{TS} \equiv 2 \ln \left( \frac{\max_{n_s}\mathcal{L}(n_s)}{\mathcal{L}(n_s = 0)} \right) = 2 \ln \left( \frac{\mathcal{L}(\hat{n}_s)}{\mathcal{L}(n_s = 0)} \right),
\end{equation}
and is a quantity which depends on the choice of model.

To account for the look-elsewhere effect, a combined background TS distribution is constructed under the null hypothesis using an ensemble of pseudo-experiments. For each pseudo-experiment, the RA of all events is randomized and the TS calculated for all of the four different models. The highest TS value out of the four is recorded and enters into the combined-background distribution, as shown by the blue error bars in Extended Data Fig.~\ref{fig:bkg_TS}. Similarly, we select the highest TS value out of the four models tested in data. The result is indicated by the black dotted line. The final ensemble consists of about 215 million pseudo-experiments. The background-TS distribution is fitted with a chi-squared distribution, which is then used to obtain the global p-value of the most-significant model. Note that the background distributions and data TS value for a specific model are also obtained in this manner, but skipping the last step of choosing the maximum over the four models.

\subsection*{Interpretation of $\hat{n}_s$ as the model normalization}

In order to interpret the result in terms of the nominal flux prediction, $\Phi_\mathrm{model}$, the model normalization is given by $\lambda(\hat{n}_s) / \lambda_0$,
where $\lambda_0$ is calculated by summing Eq.~\eqref{eq:lambdas}, with $\Phi=\Phi_\mathrm{model}$, over all datasets, and $\lambda(\hat{n}_s)$ is a function that maps $\hat{n}_s$ to the corresponding number of expected signal events. The function $\lambda$ reduces $\hat{n}_s$ bias, which can be caused by simplifications in the likelihood (such as the signal-PDF factorisation), and is determined from injection-recovery pseudo-experiments that scan over a large number of possible flux normalizations. For each pseudo-experiment, corresponding to an assumed normalization, we first calculate the expected number of signal events with Eq.~\eqref{eq:lambdas}. Then, for each dataset, MC is used to distribute $\lambda_s^j$ signal events assuming the full template shape, without energy-direction factorisation. Backgrounds are constructed from RA-randomized data. Finally, the same procedure as in data is performed to find $\hat{n}_s$, corresponding to the parameter value that maximizes Eq.~\eqref{eq:llh} for a pseudo-experiment with a total of $\lambda_s$ injected signal events. Because events are randomly sampled, each $\lambda_s$ maps to multiple $\hat{n}_s$ values; the median over all $\hat{n}_s$ for each $\lambda_s$ is used to construct $\lambda(\hat{n}_s)$.

Extended Data Fig.~\ref{fig:nsbias} shows $\lambda(\hat{n}_s)$ for the Fermi-LAT $\pi^0$ (left), both KRA$_\gamma$ (center) and CRINGE (right) templates. For each $\lambda_s$, the coloured band indicates the 16th and 84th percentiles of the resulting $\hat{n}_s$, with the solid line corresponding to the median. The correction function $\lambda(\hat{n}_s)$ is then obtained by setting $\lambda^{-1}(\lambda_s) \equiv \mathrm{Median}(\hat{n}_s)$. Inverting this relationship and treating $\mathrm{Median}(\hat{n}_s)$ as an independent variable yields $\lambda(\hat{n}_s)$, as indicated by the solid lines. The best-fit $\hat{n}_s$ from data for each template is shown as the dashed line.

Thus, although the factorisation in Eq.~\eqref{eq:sxe} is an approximation that enters the likelihood, dedicated injection-recovery tests using the full and in general angular-dependent model spectra confirm that $\lambda(\hat{n}_s)$ is unbiased under this approximation; the complete signal PDF is used for the signal injection step. However, in the presence of angular-dependent spectra, use of the factorized signal PDF means that correlations between directional and energy terms are lost, resulting in a suboptimal likelihood. The primary effect is a potential loss of statistical power rather than a systematic shift in the inferred flux.

\subsection*{Expected significance from pseudo-experiments}

After analysing the data and obtaining $\lambda(\hat{n}_s) / \lambda_0$ for all four tested models, injection-recovery tests were performed to check the consistency of the obtained significances against pseudo-experiments. Here, a pseudo-experiment corresponds to a random sampling of signal events for a given model, where the total number of events is set to $\lambda(\hat{n}_s)$, and directions and energies are sampled from the full signal PDF. It is reasonable to keep the total event number fixed as its fluctuations have a much smaller impact than the directional randomization. Backgrounds are constructed from RA-randomized data. For each pseudo-experiment, an identical procedure as in data was performed, resulting in four differing TS values that correspond to the four tested models. Comparison of the TS against the background-only TS distribution (cf.~Extended Data Fig.~\ref{fig:bkg_TS}) yields the significance.

The results of both data and the injection-recovery study are shown in Extended Data Fig.~\ref{fig:roundtrip}. Black data points show the significances obtained for data. The error bars are derived from 4000 pseudo-experiments where the signal was drawn according to the model indicated on the horizontal axis, scaled by $\lambda(\hat{n}_s) / \lambda_0$, and encompass 16th to 84th percentile with the centre point corresponding to the median. The behaviour of data significances across all four tested templates is more in line with signal drawn from the Fermi-LAT $\pi^0$ or CRINGE models than those drawn from the KRA$_\gamma$ models. This can be seen in the left and right panels where the data significance lies above that expected from pseudo-experiments drawn from the two KRA models.

\subsection*{Sensitive energy calculation}

The sensitive energy range of all four models is determined based on the energy-dependent TS distribution of the data. The 68\% (90\%) sensitive energy range is defined as the smallest possible energy interval containing 68\% (90\%) of the total TS. In order to evaluate the TS on a restricted energy interval, both the directional PDF, $\mathcal{N}^j$, and energy PDF, $\mathcal{E}^j$, are modified by restricting the integrals over true energy, $E$, in Eqs.~\eqref{eq:spdf} and \eqref{eq:epdf} to some interval that is a subset of its full range. The modified signal PDF enters into Eq.~\eqref{eq:llh}, and the TS over the restricted interval is given by $2 \ln \left( \mathcal{L}(\hat{n}_s  \lambda_0' / \lambda_0 ) /  \mathcal{L}(0) \right)$, where $\lambda_0'$ corresponds to expected number of signal events within the limited energy range, assuming $\Phi=\Phi_\mathrm{model}$. Note that $\hat{n}_s$ is still obtained from the fit over the full energy range. That is, no likelihood maximization is performed over the restricted energy interval.

Since this is a data-driven approach, correlations in the curves of the different templates in the left panel of Extended Data Fig.~\ref{fig:intervals}  are expected, as they will share the same high TS events near the Galactic Centre (GC). The harder spectrum of the KRA$_\gamma$ templates leads to higher sensitive energy ranges.

\subsection*{Flux uncertainty calculation and treatment of systematics}
The construction of the uncertainty bands of the flux measurement follows the Feldman Cousins (FC) method~\autocite{Feldman:1997qc}. Dedicated systematic MC datasets accounting for variations in optical efficiency, hole ice, and bulk ice properties are used, in order to incorporate detector-related systematic uncertainties in the model normalization~\autocite{IceCube:2019lxi}. The approach is based on a comparison of the profile likelihood of the measurement to the 68th percentile of $2 \ln \left( \mathcal{L}(\hat{n}_s) / \mathcal{L}(\lambda^{-1}(\lambda_s)) \right)$, evaluated over an ensemble of pseudo-experiments across different flux normalizations. As before, $\hat{n}_s$ corresponds to the best-fit number of signal events for a pseudo-experiment with $\lambda_s$ injected signal events, which depends on the flux normalization, and $\lambda$ is the correction function. When converting the flux to the injected $\lambda_s$, MC sets with different variations of detector systematic uncertainties are used.

To illustrate, the profile likelihood across different Fermi-LAT $\pi^0$ model normalizations is shown as the black line in the right panel of Extended Data Fig.~\ref{fig:intervals}. The colour-map shows the column normalized delta-log-likelihood distributions obtained from systematically varied pseudo-experiments, for each flux normalization bin. Red lines indicate different percentile levels obtained from those distributions. The flux values where the profile likelihood and the 68th percentile intersect yield the \SI{1}{\textsigma} uncertainty intervals. An independent ensemble of pseudo-experiments was used to confirm accurate coverage of the true model normalization.

\subsection*{Optimisation of the data residual in Galactic longitude}

The binning and event-selection criteria used for the Galactic longitude profiles in Fig.~\ref{fig:shower-longitude-excess} and Extended Data Fig.~\ref{fig:track-longitude-residuals} were determined a posteriori, based on the best-fit Fermi--LAT $\pi^{0}$ template obtained in the likelihood analysis. This optimisation is used solely for the purpose of visualising and validating the signal excess through a data-driven background subtraction, and does not enter the likelihood fit itself.

The reconstructed Galactic longitude, \(\hat{\ell}\), was binned in coarse \(40^\circ\) intervals to ensure sufficient statistics while remaining agnostic to small-scale structure.
Next, a region of interest based on Galactic latitude, $b_{\max}$, and energy threshold, $E_{\min}$, can be defined such that events with $\hat{b} > b_{\max}$ and $\hat{E} > E_{\min}$ are considered to be background (off-plane), while those with $\hat{b} < b_{\max}$ and $\hat{E} > E_{\min}$ are treated as a combination of signal and background (on-plane). The energy threshold removes events at lower reconstructed energies, which tend to have larger directional uncertainties, and reduces the background contamination. For a given longitude bin, $\hat{\ell}_i$, the expected background in the on-plane region was estimated based on the total number of events in the off-plane region, scaled down to match the solid angle of the on-plane region. The background-subtracted residual, $R(\hat{\ell}_i)$, was obtained by subtracting the scaled background from the total on-plane ($\hat{b}<b_{\max}$) event count in bin $\hat{\ell}_i$. As the bin can span a large range of declinations due to the transformation from galactic to equatorial coordinates, the declination-dependent detector sensitivity was accounted for by performing the calculation in small increments of 0.05 in $\sin \delta$, except near the horizon for the tracks sample, which uses a smaller increment of 0.005 to more accurately capture the steeply falling event rate of NT near the horizon. Statistical uncertainties, $\omega(\hat{\ell}_i)$, were obtained by propagating Poisson errors from the on- and off-plane regions. This method allows the on-plane distribution of $\hat{\ell}$ to serve as a robust probe of large-scale diffuse emission.

The optimisation of $b_{\max}$ and $E_{\min}$ was performed using Monte Carlo pseudo-experiments, with signal events injected assuming the best-fit normalization of the Fermi--LAT $\pi^{0}$ model. For each pseudo-experiment, the same on--off plane background subtraction procedure applied to the data was performed, including identical longitude binning and declination-dependent background estimation.

A scan was carried out over $b_{\max}$ and $E_{\min}$, and for each
configuration a figure of merit was computed as
\begin{equation}\label{eq:FoM}
\mathrm{FoM} = \sum_{\hat{\ell}_i} \frac{R(\hat{\ell}_i)}{\omega(\hat{\ell}_i)},
\end{equation}
where the sum runs over all longitude bins. The optimal selection was chosen by
maximising this figure of merit, corresponding to the configuration that yields
the largest excess over the isotropic background along the GP when
accounting for statistical uncertainties. The resulting selections, applied separately for shower-like and track-like
events to account for their different angular resolutions and energy
estimators, are reported in the main text.

%% file: authorlist.tex
R. Abbasi$^{16}$,
M. Ackermann$^{63}$,
J. Adams$^{17}$,
J. A. Aguilar$^{10}$,
M. Ahlers$^{21}$,
J.M. Alameddine$^{22}$,
S. Ali$^{35}$,
N. M. Amin$^{43}$,
K. Andeen$^{41}$,
C. Arg{\"u}elles$^{13}$
% ...
\\
\\
% ...
$^{10}$ Universit{\'e} Libre de Bruxelles, Science Faculty CP230, B-1050 Brussels, Belgium \\
% ...
$^{13}$ Department of Physics and Laboratory for Particle Physics and Cosmology, Harvard University, Cambridge, MA 02138, USA \\
% ...
$^{16}$ Department of Physics, Loyola University Chicago, Chicago, IL 60660, USA \\
$^{17}$ Dept. of Physics and Astronomy, University of Canterbury, Private Bag 4800, Christchurch, New Zealand \\
% ...
$^{21}$ Niels Bohr Institute, University of Copenhagen, DK-2100 Copenhagen, Denmark \\
$^{22}$ Dept. of Physics, TU Dortmund University, D-44221 Dortmund, Germany \\
% ...
$^{35}$ Dept. of Physics and Astronomy, University of Kansas, Lawrence, KS 66045, USA \\
% ...
$^{41}$ Department of Physics, Marquette University, Milwaukee, WI 53201, USA \\
% ...
$^{43}$ Bartol Research Institute and Dept. of Physics and Astronomy, University of Delaware, Newark, DE 19716, USA \\
% ...
$^{63}$ Deutsches Elektronen-Synchrotron DESY, Platanenallee 6, D-15738 Zeuthen, Germany \\
% ...

%% file: ed.tex
\begin{edfigure}[htbp]
\centering
    \includegraphics[width=\linewidth]{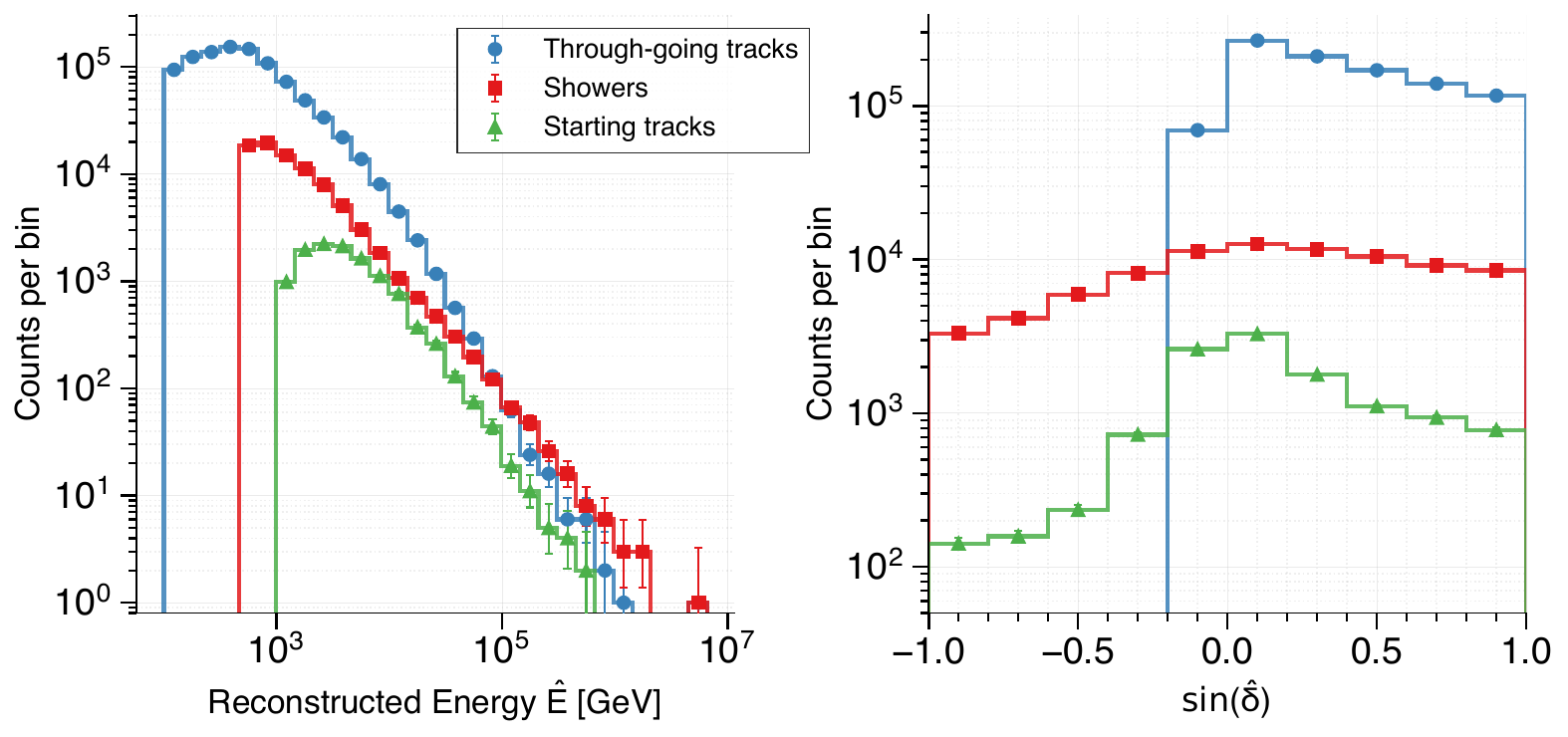}
    \caption{\textbf{Data distributions for individual samples.} We present data histograms in terms of reconstructed energy \(\hat{E}\) (left) and \(\sin\hat{\delta}\) (right) for through-going tracks (blue), showers (red), and starting tracks (green) used in this analysis. Note that different reconstruction algorithms are used for each dataset. Through-going tracks are predominantly from the northern sky, using the Earth to suppress atmospheric-muon background. Showers and starting tracks cover the full sky. Overlapping events between subsamples are removed such that they only appear once in the combined dataset, following the prescription described in the text. Error bars show 68\% Poisson (Garwood) intervals.}
    \label{fig:data_distributions}
\end{edfigure}
\vspace{-1em}

\begin{edfigure}[!tbp]
    \centering
    \includegraphics[width=0.5\linewidth]{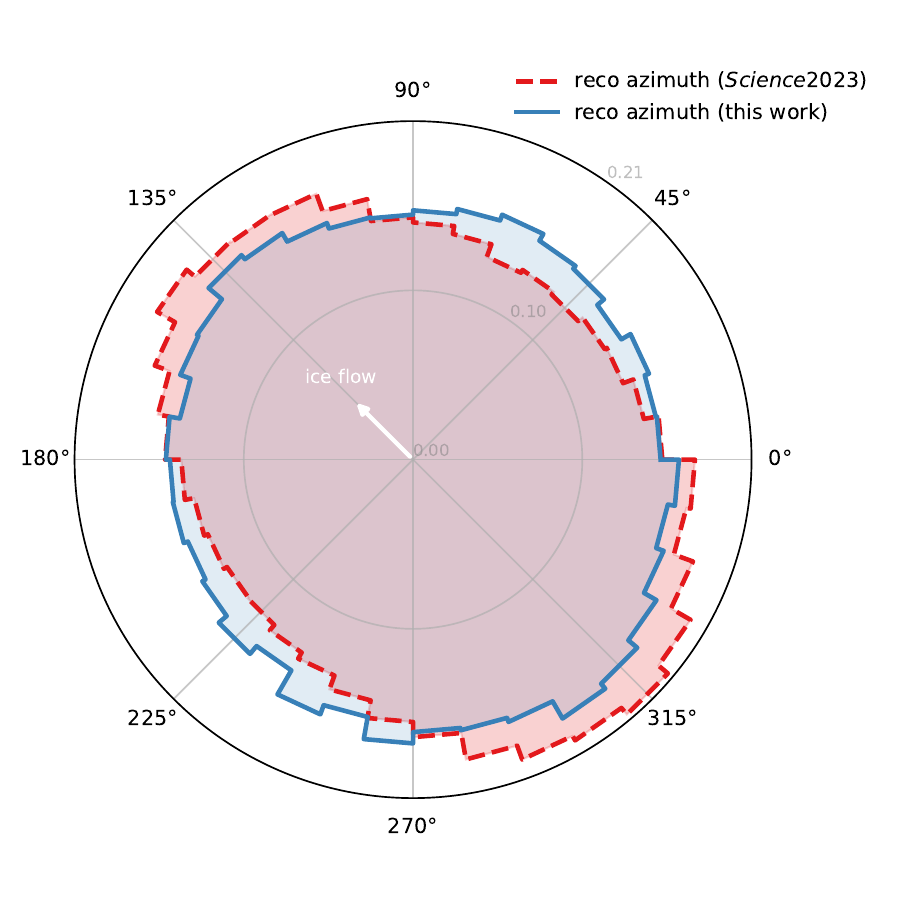}
    \vspace{-0.6em}

    \caption{
    \textbf{Azimuthal distributions of shower-like events in 12 years of IceCube data.}
    The red dashed contour corresponds to the prior reconstruction~\autocite{IceCube:2023ame}, which relied on a deep neural network trained on simulations that used a now-outdated ice model to predict the expected Cherenkov light yield at each sensor. An anisotropy aligned along the ice-flow direction is visible, indicative of artefacts induced by unmodelled ice properties. The blue solid contour corresponds to the updated maximum-likelihood reconstruction~\autocite{IceCube:2024csv} of shower-like events used in this work, which incorporates ice birefringence~\autocite{tc-18-75-2024}, layer undulations~\autocite{IceCube:2023qua}, and an updated calibration of the refrozen hole ice~\autocite{IceCube:2013llx}. The radial scale indicates the normalised probability per azimuth bin.
    }
    \label{fig:iceman_azimuth}
\end{edfigure}

\vspace{-0.4cm}

\begin{edfigure}[!tbp]
    \centering

    \begin{minipage}{0.48\linewidth}
        \centering
        \includegraphics[width=\linewidth]{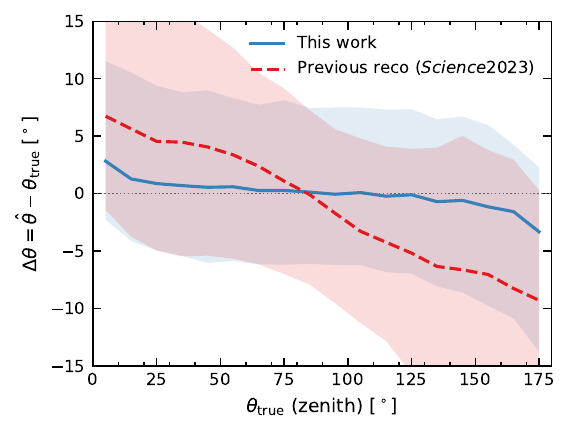}\\[-0.4em]
        \textbf{a}
    \end{minipage}
    \hfill
    \begin{minipage}{0.48\linewidth}
        \centering
        \includegraphics[width=\linewidth]{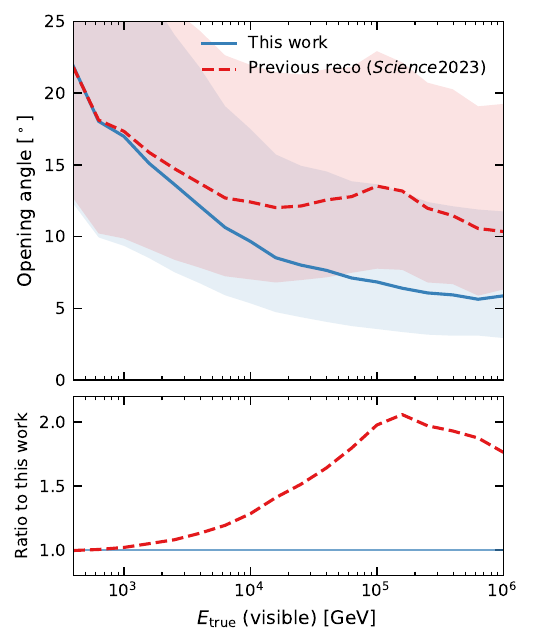}\\[-0.4em]
        \textbf{b}
    \end{minipage}

    \vspace{-0.6em}

    \caption{
    \textbf{Simulation-based comparison of shower-event reconstruction performance.} The simulation includes ice birefringence~\autocite{tc-18-75-2024}, layer undulations~\autocite{IceCube:2023qua}, and an updated calibration of the refrozen hole ice~\autocite{IceCube:2013llx}.
    \textbf{(a)} Zenith angle bias evaluated on simulation. The signed zenith angle residual is defined as $\Delta\theta=\hat{\theta}-\theta_{\mathrm{true}}$. Solid lines show the median, and shaded bands indicate the 16th--84th percentile range. The prior reconstruction, shown in red, exhibits a systematic horizon-compression bias up to $\ang{10}$: events are preferentially reconstructed toward $\hat{\theta}\approx\ang{90}$ regardless of true direction. The updated reconstruction, shown in blue, exhibits less bias across all zenith angles. Small deviations from zero at the extremes are artefacts that occur by construction for $\theta_{\mathrm{true}}=\ang{0}$ and $\ang{180}$.
    \textbf{(b)} Angular-resolution comparison using the same simulation sample. The opening angle is defined as the angular separation between the true neutrino direction and the reconstructed neutrino direction. The red dashed line shows the prior cascade reconstruction~\autocite{IceCube:2023ame}; the blue solid line shows the updated maximum-likelihood reconstruction~\autocite{IceCube:2024csv}. Shaded bands indicate the 25th--75th percentile range. The prior reconstruction exhibits a worse angular resolution over the full energy range. The upturn at around \SI{100}{\tera\eV} disappears when recent updates in calibration are included as part of the neural-network training.  The lower panel shows the ratio of the prior reconstruction's median opening angle to that of the updated reconstruction.
    }
    \label{fig:iceman_reco_sim}
\end{edfigure}

\begin{edfigure}[p]
    \centering
    \includegraphics[width=0.8\linewidth]{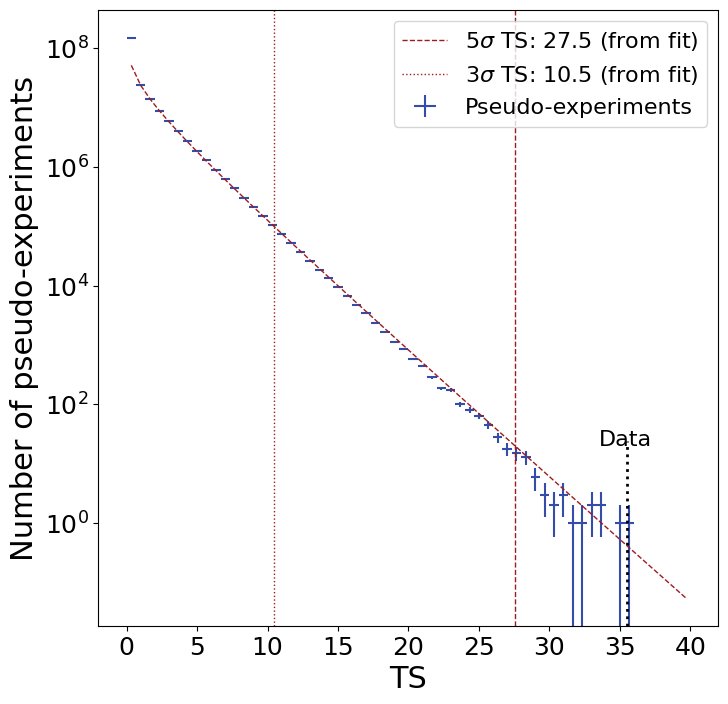}
    \caption{\textbf{Combined background-TS distribution.} The background-TS distribution, shown as the blue error bars, is constructed from 215 million pseudo-experiments. Each TS is obtained by fitting a RA-randomized dataset with all four model templates and recording the highest TS value out of the four. The resulting distribution is then fitted with a chi-square distribution, shown as the dashed, red line. The resulting best-fit corresponds to $\chi^2(k=1.16)$, where $k$ is the number of degrees of freedom, and is used to map the TS to a significance level (\SI{3}{\textsigma} and \SI{5}{\textsigma} levels are indicated by the red, vertical lines). The TS for the observed data is highlighted as the black, dashed line, corresponding to a global significance of \SI{5.7}{\textsigma}.
}
    \label{fig:bkg_TS}
\end{edfigure}

\begin{edfigure}[p]

    \begin{center}
    \begin{subedfigure}[t]{0.33\linewidth}
        \includegraphics[width=\linewidth]{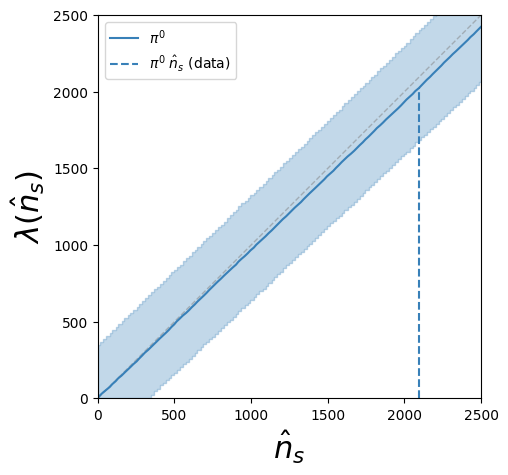}
    \end{subedfigure}
    \begin{subedfigure}[t]{0.315\linewidth}
        \includegraphics[width=\linewidth]{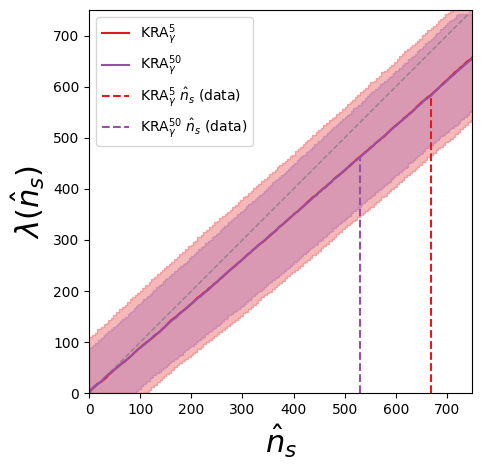}
    \end{subedfigure}
    \begin{subedfigure}[t]{0.32\linewidth}
        \includegraphics[width=\linewidth]{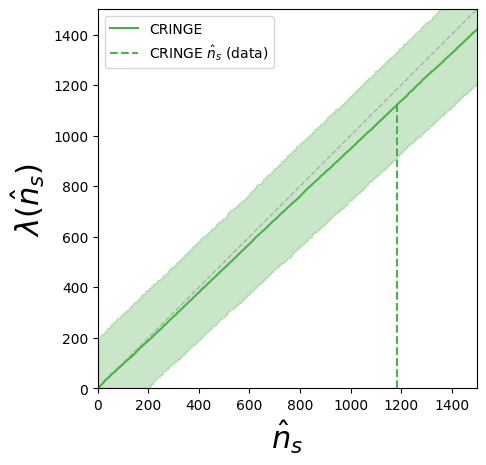}
    \end{subedfigure}
    \end{center}

    \caption{\textbf{Construction of $\lambda(\hat{n}_s)$.} The relationship between $\lambda_s$, injected from pseudo-experiments, and their corresponding $\hat{n}_s$ is shown for each of the four tested templates. For each $\lambda_s$, coloured bands correspond to the 16th and 84th percentiles of the resulting  $\hat{n}_s$ distributions, with the solid line corresponding to the median. Inverting the median and reinterpreting it as the independent variable yields $\lambda(\hat{n}_s)$, as indicated in the legend. The best-fit $\hat{n}_s$ from data is shown as dashed, vertical lines for each of the different templates.
    }

    \label{fig:nsbias}
\end{edfigure}

\begin{edfigure}[p]
    \centering
    \includegraphics[width=\linewidth]{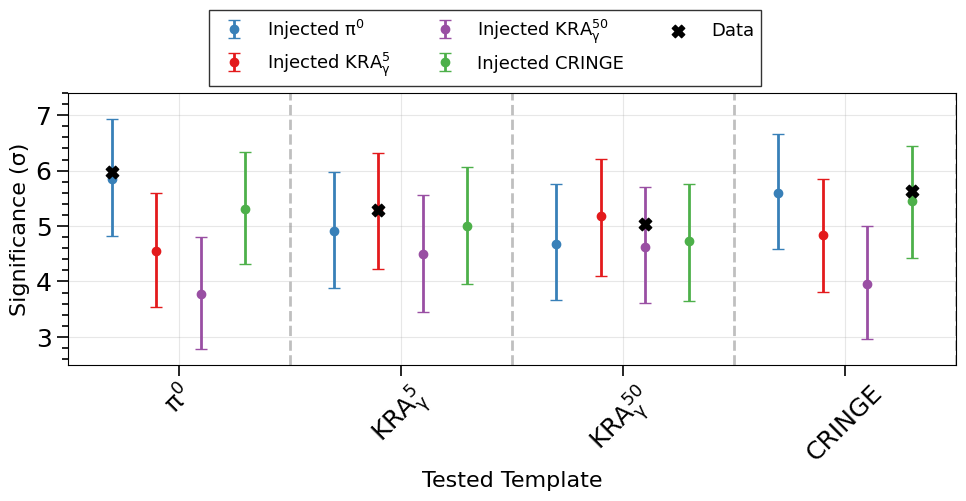}
    \caption{\textbf{Comparison of expected significances to data.} Comparison of local significances obtained for data for each of the four tested models (black) against pseudo-experiments (error bars) where signal events are drawn according to the model indicated on the horizontal axis. The error bars encompass the 16th to 84th percentiles from 4000 pseudo-experiments, each calculated according to the prescription described in the text. The data exhibits better agreement with signal drawn from the Fermi-LAT $\pi^0$ or CRINGE models than the KRA$_\gamma$ models.}
    \label{fig:roundtrip}
\end{edfigure}

\begin{edfigure}[p]
    \centering
    \includegraphics[width=0.42\linewidth]{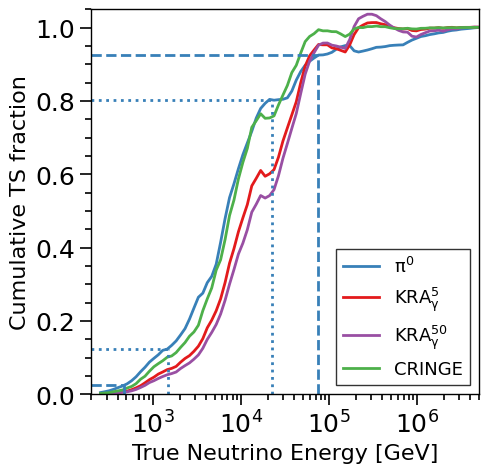}
    \includegraphics[width=0.55\linewidth]{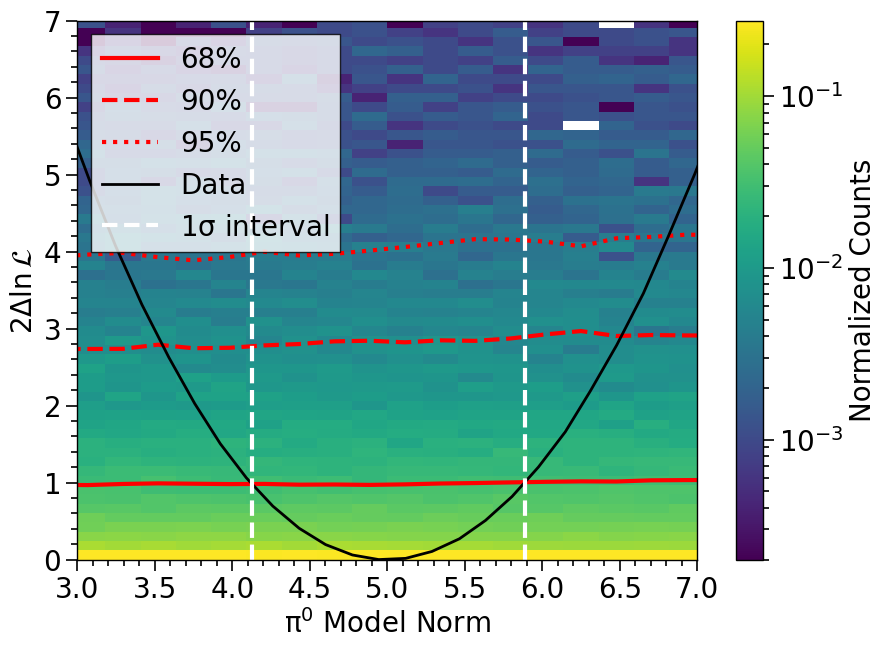}
    \caption{\textbf{Sensitive energy range and uncertainty interval construction.} The left panel shows the fraction of the best fit TS that remains when cutting the MC, used for the calculation of the signal PDF $\mathcal{S}^j$, at an upper bound shown on the x-axis. At each step in energy, the cumulative TS is calculated as $2 \ln (\mathcal{L}(\hat{n}_s  \lambda_0'/ \lambda_0) / \mathcal{L}(0))$, where $\hat{n}_s$ is obtained via Eq.~\eqref{eq:TS} over the full energy range and $\lambda_0' / \lambda_0$ corresponds to the expected fraction of signal events below the energy upper bound. Based on the cumulative TS, the 68\% (90\%) sensitive energy range can be obtained, and as an example the Fermi-LAT $\pi^0$ is model is shown as the dotted (dashed) lines. The right panel illustrates the normalization uncertainty band construction using the Feldman-Cousins approach~\autocite{Feldman:1997qc}. Each column corresponds to 5,000 pseudo-experiments over a fixed range of \textit{Fermi}-LAT $\pi^0$ model normalisations. For each column, the colour-map shows the probability per bin of $2 \Delta \ln \mathcal{L} \equiv 2 \ln \left( \mathcal{L}(\hat{n}_s) / \mathcal{L}(\lambda^{-1}(\lambda_s)) \right)$, of which the 68th percentile is indicated by the solid red line. The intersection of this threshold and the profile likelihood of the data fit (black curve) is the reported \SI{1}{\textsigma} interval (white dashed lines).
}
    \label{fig:intervals}
\end{edfigure}

\begin{edfigure}[p]
  \centering
  \includegraphics[width=0.95\textwidth]{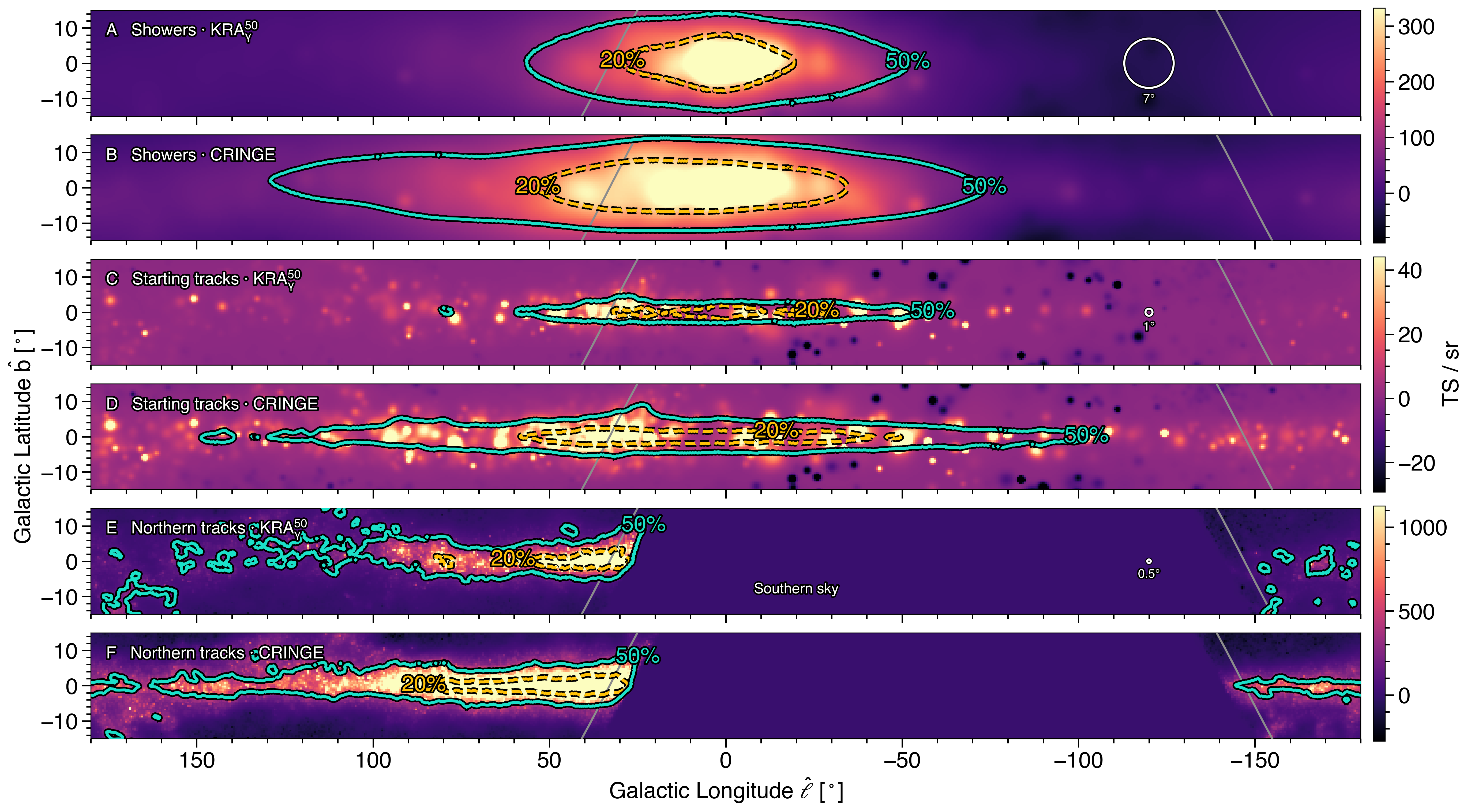}
  \caption{\textbf{Test statistic (TS) distributions for CRINGE and $\mathrm{KRA}_{\gamma}^{50}$ templates.} Same as Fig.~\ref{fig:iceman_TS_map} but for CRINGE (panels A, C, E) and $\mathrm{KRA}_{\gamma}^{50}$ (panels B, D, F). Colour maps show the TS per steradian for the shower (A, B), starting-track (C, D) and northern-track (E, F) sample, as indicated by the colour bar (unique to each sample). Contour lines indicate \SI{20}{\%} and \SI{50}{\%} containment regions of the signal
PDF for the corresponding template, assuming a typical angular uncertainty for an event from the
sample (white circle, 7° for showers, 1° for starting tracks and 0.5° for through-going tracks). Grey lines mark $\delta = \ang{0}$ in equatorial coordinates.}
  \label{fig:ts_stack_method}
\end{edfigure}


\begin{edfigure}[p]
\centering
\includegraphics[width=0.9\textwidth]{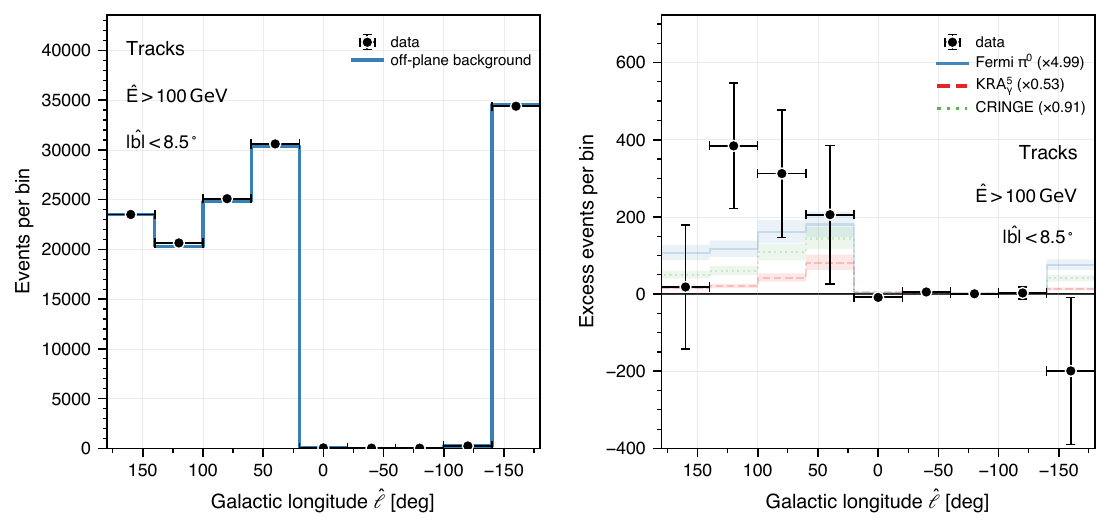}
\caption{
\textbf{Distribution and background-subtracted residuals of track events along the Galactic longitude.} The left panel shows observed counts (black points) for track events with reconstructed energy $\hat{E}>\SI{0.1}{\tera \eV}$ and Galactic latitude $|\hat{b}|<\ang{8.5}$, with error bars corresponding to the uncertainty from Poisson statistics.
The blue histogram shows data-derived background expectations, estimated from off-plane regions along the same declination; the shaded band corresponds to its statistical uncertainty. The right panel shows residuals (black points) after background subtraction, $N_{\mathrm{obs}}(\hat{\ell})-N_{\mathrm{bkg}}(\hat{\ell})$. The uncertainties include the Poisson uncertainty of the observed counts and the propagated statistical uncertainty of the scaled off-plane background estimate. 
Coloured lines show signal expectations from three of the tested templates, scaled to their best-fit normalisations from the likelihood analysis; shaded bands indicate the corresponding \SI{1}{\textsigma} uncertainty on the normalisation. Compared to Fig.~\ref{fig:shower-longitude-excess}, a weaker excess is visible in the northern sky. Model expectations are overlaid for visual comparison only; no fit is performed to the binned points shown here.
}
\label{fig:track-longitude-residuals}
\end{edfigure}


\begin{edfigure}[p]
    \centering
    \includegraphics[width=0.8\linewidth]{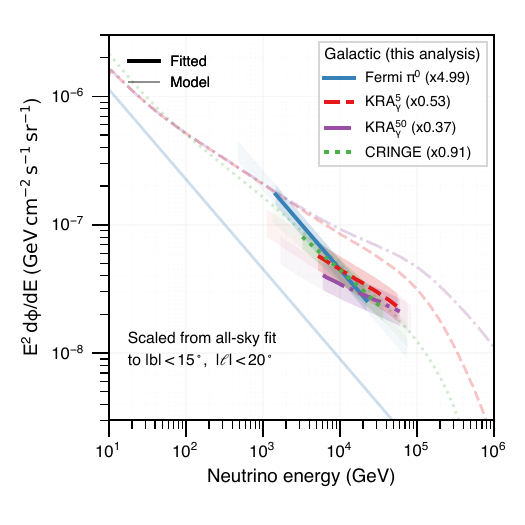}

   \caption{\textbf{Diffuse Galactic neutrino spectra and model comparison in the inner Galaxy.}  The best-fit all-flavour differential flux for four Galactic emission models are shown in: \textit{Fermi}-LAT~$\pi^0$ (blue), KRA$_\gamma^{5}$ (red), KRA$_\gamma^{50}$ (purple), and CRINGE (green). Thin lines indicate the nominal model predictions, while thick lines show the spectra scaled by the best-fit normalisation obtained from the global likelihood analysis. The spectra are evaluated over the inner Galactic region \((|b|\le15^\circ,\, |\ell|\le20^\circ)\). The energy range contributing most strongly to the fit is indicated by shaded bands (central 68\% in darker shading and 90\% in lighter shading). Vertical error bars denote 68\% confidence intervals derived using a unified Feldman--Cousins construction~\autocite{Feldman:1997qc}, including detector systematic uncertainties.
    }
      
 \label{fig:model_fits}
\end{edfigure}

%% file: supplementary.tex
\vspace*{1em}
\begin{center}
{\LARGE Supplementary Information for ``High-energy neutrino emission from the Milky Way''\par} 
\vspace{1.5em} {\large The IceCube Collaboration\par}
\end{center} 
\setcounter{page}{1}
\begin{refsection}
\section{Effective areas and detector acceptance}
The effective area $A_{\rm eff}(\bOmega, E)$ quantifies the detection efficiency of the detector for an incident neutrino flux as a function of direction $\bOmega$ and true neutrino energy $E$, and has units of area. It encapsulates the combined effects of the neutrino interaction cross section, absorption in the Earth, detector geometry, event selection criteria, and reconstruction efficiencies. Supplementary Fig.~\ref{fig:eff_area} shows the effective area as a function of energy for different declination bands, event topologies, and neutrino flavours.

\begin{sfigure}[htbp]
    \centering
    \begin{subsfigure}[t]{0.45\linewidth}
        \centering
        \includegraphics[width=\linewidth]{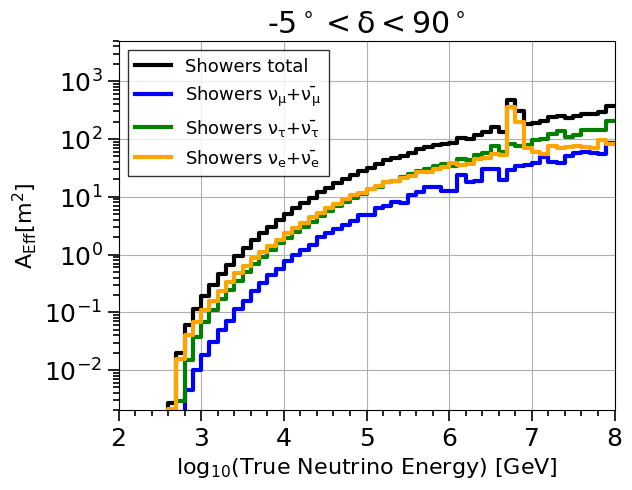}
        \caption{}
    \end{subsfigure}
    \begin{subsfigure}[t]{0.45\linewidth}
        \centering
        \includegraphics[width=\linewidth]{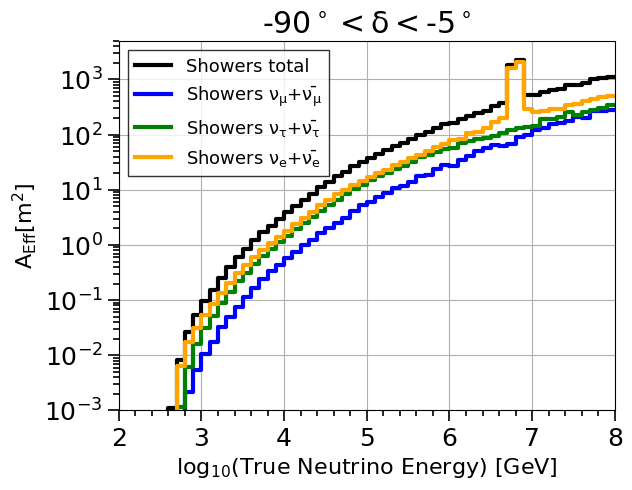}
        \caption{}
    \end{subsfigure}
    \begin{subsfigure}[t]{0.45\linewidth}
        \centering
        \includegraphics[width=\linewidth]{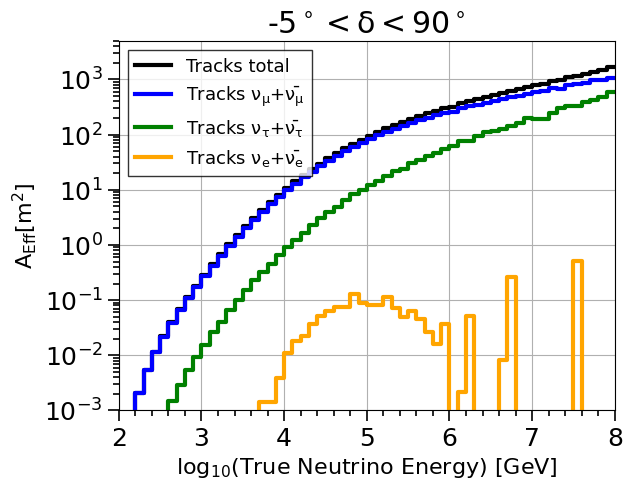}
        \caption{}
    \end{subsfigure}
    \begin{subsfigure}[t]{0.45\linewidth}
        \centering
        \includegraphics[width=\linewidth]{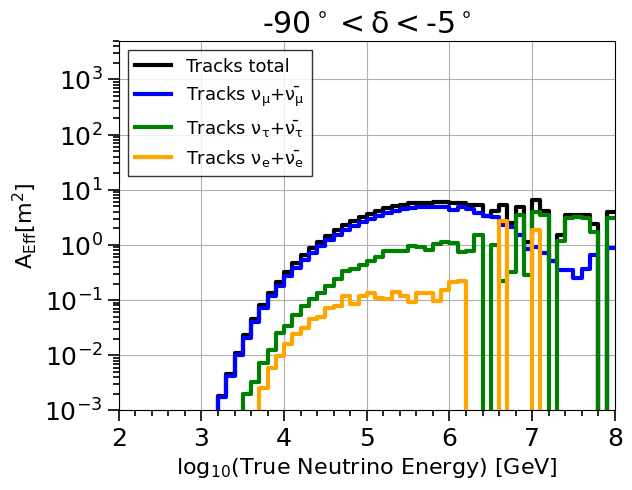}
        \caption{}
    \end{subsfigure}
    \begin{subsfigure}[t]{0.45\linewidth}
        \centering
        \includegraphics[width=\linewidth]{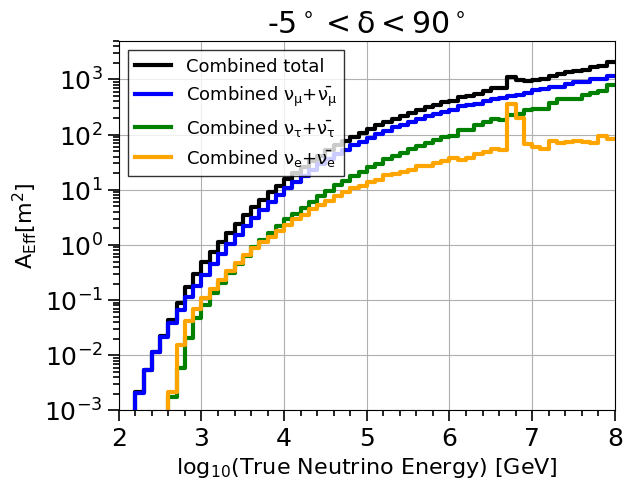}
        \caption{}
    \end{subsfigure}
    \begin{subsfigure}[t]{0.45\linewidth}
        \centering
        \includegraphics[width=\linewidth]{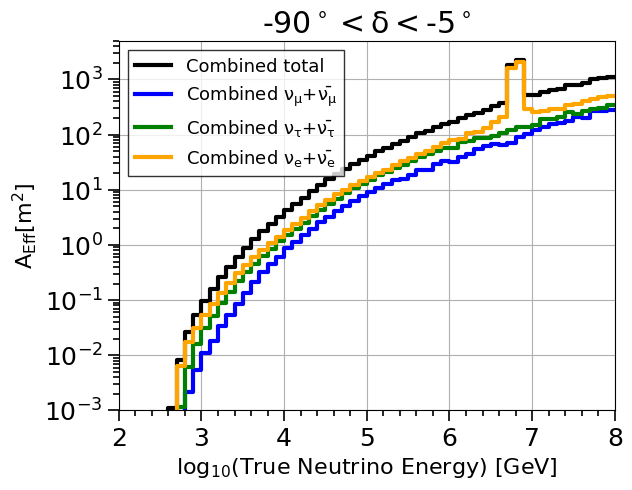}
        \caption{}
    \end{subsfigure}

    \caption{\textbf{Neutrino effective areas per flavour.} Results of MC simulations are shown in terms of the true neutrino energy for the northern (left column) and southern (right column) sky and presented for shower events (DNNC in (a) and (b)), track events (NT+ESTES in (c) and (d)), as well as the combined sample ((e) and (f)).}

    \label{fig:eff_area}
\end{sfigure}
For the comparisons shown in Supplementary Fig.~\ref{fig:eff_area} we split the sky into southern ($-90^\circ \le \delta < -5^\circ$) and northern ($-5^\circ \le \delta \le 90^\circ$) declination bands and present results for showers, tracks, and the combined sample. At the highest energies, Earth absorption suppresses neutrinos arriving from below the horizon; features due to the Glashow resonance (around \SI{6.3}{\peta \eV}) are most visible in the southern-sky shower sample. In the northern sky near 100\,TeV, the muon-neutrino effective area exceeds that for electron neutrinos by nearly an order of magnitude, with tau neutrinos in between. This hierarchy reflects the larger effective volume for through-going tracks: muons can traverse longer distances, enhancing acceptance, whereas electrons interact promptly to produce compact showers. Tau-neutrino acceptance lies between the two because tau leptons originating from charged-current interactions have multiple decay modes; at $\sim$100\,TeV the mean decay length is of order a few metres, yielding mostly shower-like topologies unless the tau leptonically decays to a muon (branching ratio $\approx17\%$).

\begin{sfigure}[htbp]
    \centering
    \begin{subsfigure}[t]{\linewidth}
        \centering
        \includegraphics[width=\linewidth]{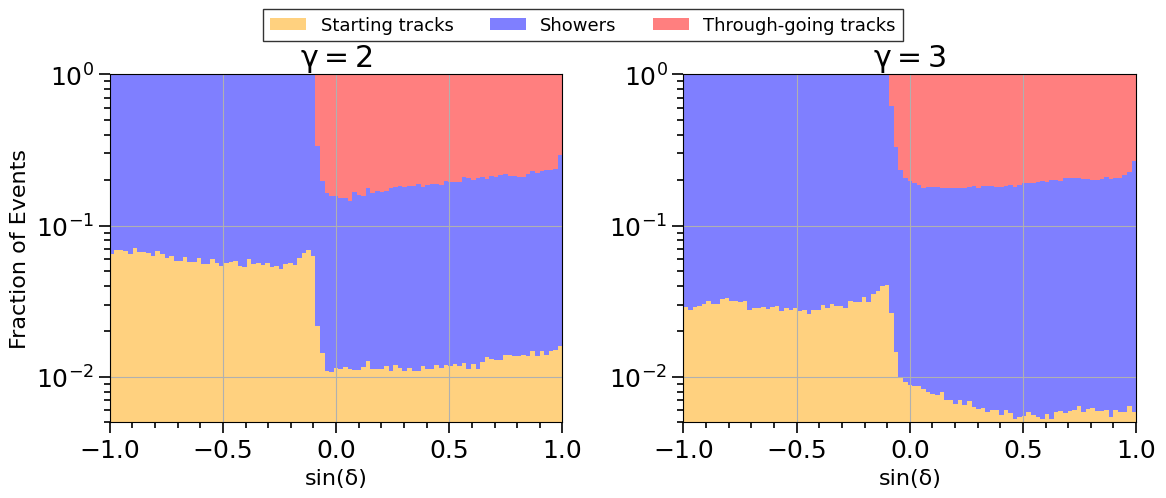}
        \caption{}
    \end{subsfigure}
    
    \caption{\textbf{Expected contributions of each dataset.} Fractional event counts integrated over energy assuming power-law neutrino spectra following $E^{-2}$ (left) and $E^{-3}$ (right), shown as a function of declination for the three event classes. Through-going muon tracks dominate the event rate in the northern sky, while shower-like events dominate in the Southern sky.}
    \label{fig:eff_area_2d}
\end{sfigure}
Another visualization of the relative contributions of the different event samples as a function of declination is shown in Supplementary Fig.~\ref{fig:eff_area_2d}. The vertical axis shows the integrated fraction of Monte-Carlo (MC) events expected in each sample over the full energy range, assuming a power-law neutrino flux following $E^{-2}$ (left) and $E^{-3}$ (right). This representation illustrates how the detector response redistributes sensitivity across event classes and sky regions.

In the northern sky, the sample is dominated by through-going muon tracks, reflecting the large effective area for up-going muon neutrinos. In the southern sky, the contribution from shower events becomes dominant, followed by starting tracks. This figure is intended to illustrate detector acceptance trends only; in the actual analysis, signal and background have different energy spectra and energy-dependent purities, which are not captured by this simplified representation.

\section{Cross checks between different datasets}
Shower-like and track-like events probe different regions of the sky and different energy ranges, and achieve maximal sensitivity under different conditions. As illustrated in Fig.~3 of the main text and described in the Methods, track-like events have highest signal-to-background ratios primarily in the northern sky and at lower reconstructed energies, below $\sim1\,\mathrm{TeV}$. In contrast, the shower sample is most sensitive to emission from the inner Galaxy at higher energies, above $\sim5\,\mathrm{TeV}$, where the atmospheric background is suppressed by the self-veto effect and the Galactic signal is expected to peak.

\begin{table}[!htbp]
  \centering
  \caption{\textbf{Flux measurements and local significances.} Best-fit normalizations (68\% C.I.) and local significances by model template and sample. The quoted uncertainties for DNNC and NT include detector systematics; ESTES reports statistical uncertainties only as its systematic simulation set does not exist. Its lower significance implies that it does not have strong constraining power on the flux. For stand-alone fits, overlaps between subsamples are not removed; overlaps are removed in the combined fit. The flux is given relative to that of the nominal model, $\Phi_\mathrm{model}$. To aid comparison with Supplement, Tab.~\ref{tab:GP_Results}, $E^2 \Phi_\mathrm{model}=\SI{4.4e-12}{\tera \eV \cm^{-2} s^{-1}}$ at \SI{100}{\tera \eV} for \textit{Fermi} $\pi^0$.}
  \label{tab:bestfit_iceman}
  \begin{threeparttable}
  \begin{tabular}{l c c c}
    \toprule
    \textbf{Template} & \textbf{Sample} & \textbf{Flux/Model norm} & \textbf{Local significance} \\
    \midrule
    \multirow{4}{*}{\textit{Fermi} $\pi^0$}
      & Combined & \asym{4.99}{0.90}{0.86} $\times$ $\Phi_\mathrm{model}$ & \SI{5.97}{\textsigma} \\
      & DNNC     & \asym{5.21}{1.10}{1.06} $\times$ $\Phi_\mathrm{model}$ & \SI{5.29}{\textsigma} \\
      & NT       & \asym{4.07}{1.82}{1.81}  $\times$ $\Phi_\mathrm{model}$ & \SI{2.22}{\textsigma} \\
      & ESTES\tnote{*}    & \asym{6.23}{4.48}{3.80} $\times$ $\Phi_\mathrm{model}$ & \SI{1.61}{\textsigma} \\
    \midrule
    \multirow{4}{*}{KRA$_\gamma^{5}$}
      & Combined & \asym{0.53}{0.14}{0.12} $\times$ $\Phi_\mathrm{model}$ & \SI{5.34}{\textsigma} \\
      & DNNC     & \asym{0.53}{0.16}{0.14} $\times$ $\Phi_\mathrm{model}$& \SI{5.05}{\textsigma} \\
      & NT       & \asym{0.60}{0.37}{0.36} $\times$ $\Phi_\mathrm{model}$& \SI{1.60}{\textsigma} \\
      & ESTES\tnote{*}    & \asym{0.33}{0.37}{0.22} $\times$ $\Phi_\mathrm{model}$ & \SI{1.23}{\textsigma} \\
    \midrule
    \multirow{4}{*}{KRA$_\gamma^{50}$}
      & Combined & \asym{0.37}{0.10}{0.09} $\times$ $\Phi_\mathrm{model}$& \SI{5.10}{\textsigma} \\
      & DNNC     & \asym{0.39}{0.12}{0.10} $\times$ $\Phi_\mathrm{model}$& \SI{4.97}{\textsigma} \\
      & NT       & \asym{0.31}{0.30}{0.22} $\times$ $\Phi_\mathrm{model}$& \SI{1.13}{\textsigma} \\
      & ESTES\tnote{*}   & \asym{0.25}{0.29}{0.16} $\times$ $\Phi_\mathrm{model}$& \SI{1.23}{\textsigma} \\
    \midrule
    \multirow{4}{*}{CRINGE}
      & Combined & \asym{0.91}{0.18}{0.18} $\times$ $\Phi_\mathrm{model}$& \SI{5.64}{\textsigma} \\
      & DNNC     & \asym{0.89}{0.21}{0.20} $\times$ $\Phi_\mathrm{model}$& \SI{4.89}{\textsigma} \\
      & NT       & \asym{0.87}{0.42}{0.39} $\times$ $\Phi_\mathrm{model}$& \SI{2.20}{\textsigma} \\
      & ESTES\tnote{*}   & \asym{1.22}{0.86}{0.71} $\times$ $\Phi_\mathrm{model}$& \SI{1.77}{\textsigma} \\
    \bottomrule
  \end{tabular}
  \begin{tablenotes}
    \item[*] Statistical uncertainties only
  \end{tablenotes}
  \end{threeparttable}
\end{table}

Table~\ref{tab:bestfit_iceman} summarises the best-fit normalizations and corresponding \SI{1}{\textsigma} confidence intervals for each of the tested Galactic templates, obtained separately from the combined all-flavour sample as well as from the individual shower (DNNC), through-going track (NT), and starting-track (ESTES) samples. For the stand-alone fits to individual event classes, overlaps between samples are not removed, whereas overlaps are removed in the combined fit. For DNNC and NT, the detector-related systematic uncertainties are included in the quoted confidence intervals, which are derived using the Feldman--Cousins approach as described in the Methods. For ESTES only statistical uncertainties are reported, as its systematic simulation set does not exist. Its lower significance implies that it does not have strong constraining power on the flux. Within uncertainties, the best-fit normalizations obtained from the different event samples are mutually consistent at the \SI{1}{\textsigma} level.

In addition, local significances are provided for each sample, based on their individually maximized test statistics (TS). As described in the Methods, the TS depends on the reconstructed energy of individual events, their directional correlation with the assumed Galactic emission templates, and the ratio of signal and background probability density functions (PDF)s that enter into the likelihood.

For the Combined, DNNC, and NT samples, the \textit{Fermi}-LAT $\pi^{0}$ template~\autocite{Fermi-LAT:2012edv} yields the most significant best-fit normalization; the combined-sample analysis yields a local significance of \SI{5.97}{\textsigma} for this template. After accounting for the look-elsewhere effect associated with additionally testing KRA$_\gamma$ and CRINGE Galactic emission models~\autocite{Gaggero:2015xza,Schwefer:2022zly}, the corresponding global significance is \SI{5.7}{\textsigma}, placing the result well beyond the conventional \SI{5}{\textsigma} threshold for discovery.

\begin{sfigure}[htbp]
    \centering
    \begin{subsfigure}[t]{\linewidth}
        \centering
        \includegraphics[width=\linewidth]{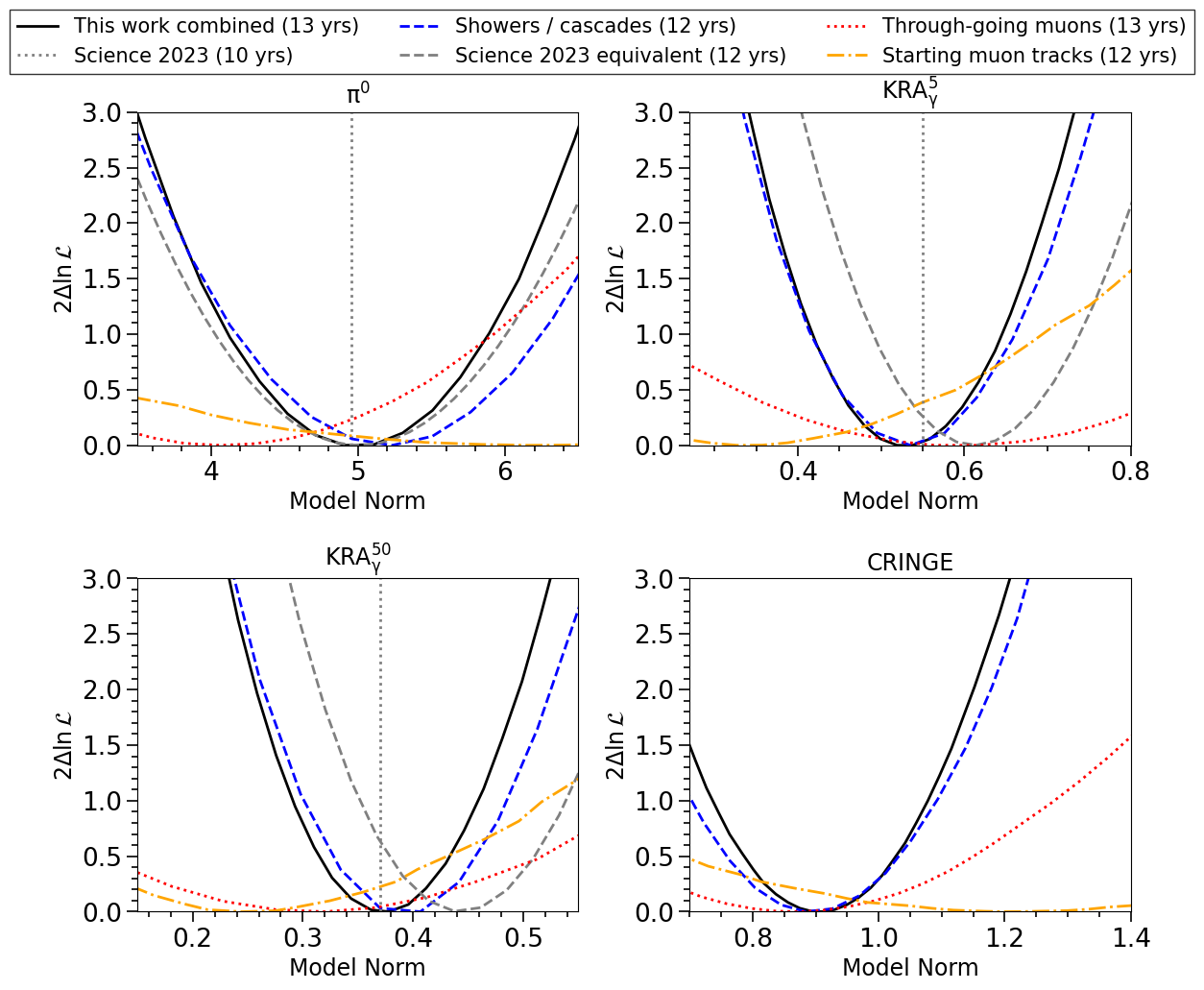}
        \caption{}
    \end{subsfigure}
    
    \caption{\textbf{Likelihood as a function of model normalisation.} Shown are the $-2\Delta\ln\mathcal{L}$ contours for all four tested Galactic emission templates, obtained from the individual event samples as well as from the combined analysis. The previous IceCube result based on ten years of shower data is shown as the vertical dotted line. As a cross-check, a 12-year shower-only analysis using the same reconstruction and simulation in the previous GP analysis~\autocite{IceCube:2023ame}, but without including systematics or testing the CRINGE model, is shown as the dashed grey line. All results are consistent within \SI{1}{\textsigma} uncertainties.}

    \label{fig:LLH}
\end{sfigure}
Likelihood contours over model normalization are shown in Supplementary Fig.~\ref{fig:LLH}. The solid black curve corresponds to the combined all-flavour fit, while the dashed blue, dotted red, and dash-dotted yellow curves show results from the DNNC, NT, and ESTES samples, respectively. The vertical dotted line indicates the best-fit values from the previous GP analysis~\autocite{IceCube:2023ame}, obtained prior to improvements in ice modelling~\autocite{tc-18-75-2024,IceCube:2023qua} and event reconstruction~\autocite{IceCube:2024csv}. Within uncertainties, no significant tension is observed between the results obtained from the different event samples. The CRINGE model, which includes an unresolved-source component rather than purely diffuse emission, yields a best-fit normalization closest to its nominal prediction. In contrast, the \textit{Fermi}-LAT $\pi^{0}$ template requires a normalization around a factor of five higher than that of the nominal prediction, indicative of a mismatch between the nominal model prediction and the observed neutrino flux, particularly in the inner Galaxy. At present, however, the data cannot conclusively determine a single, preferred template from the tested Galactic emission models.

Table~\ref{tab:sensitivity_breakdown} provides a breakdown of the \SI{5}{\textsigma} discovery potential (DP) progression, evaluated from pseudo-experiments using the same simulation and analysis chain as that used for data. As a point of comparison, the DNNC 10-year column is obtained assuming the livetime of the previous Galactic plane (GP) analysis~\autocite{IceCube:2023ame}. The impact of an additional two years of livetime and inclusion of both track samples are provided, with relative improvements (lower is better) in the \SI{5}{\textsigma} DP given in parentheses. 

\begin{table}[htbp]
\centering
\caption{\textbf{\SI{5}{\textsigma} discovery potential (DP) breakdown.}
The median flux required for a \SI{5}{\textsigma} detection, evaluated from pseudo-experiments using the same simulation and analysis chain as the results shown in Supplement, Tab.~\ref{tab:bestfit_iceman}. The DNNC 10-year column assumes the livetime of the previous Galactic plane (GP) analysis~\autocite{IceCube:2023ame} and uses the shower-only sample, the DNNC 12-year column incorporates two additional years and the Combined 12-year column adds the NT and ESTES
track samples. Values are given in units of $\Phi_\mathrm{model}$ for each template. Parenthetical values give the percentage improvement in \SI{5}{\textsigma} DP relative to the column to its left; a smaller required flux indicates better
sensitivity. The total gain is computed relative to the DNNC 10-year column. All values are rounded to the nearest percent.}
\label{tab:sensitivity_breakdown}
\begin{tabular}{lcccc}
\toprule
Template & DNNC 10yr & DNNC 12yr & Combined 12yr  & Total gain \\
\midrule
\textit{Fermi}-LAT $\pi^0$ & 5.53  & 5.11 (8\%) & 4.25 (17\%) & 23\% \\
KRA$_\gamma^5$             & 0.60  & 0.55 (8\%) & 0.49 (11\%) & 18\% \\
KRA$_\gamma^{50}$          & 0.44  & 0.40 (9\%) & 0.37 (8\%)  & 16\% \\
CRINGE                     & 1.03  & 1.00 (3\%) & 0.84 (16\%) & 18\% \\
\bottomrule
\end{tabular}
\end{table}

A search for neutrino emission from the Galactic Plane was also performed using the same shower event selection, reconstruction, and MC as in the previous GP analysis\autocite{IceCube:2023ame}, with two additional years of data. This study does not include the changes to reconstruction and calibration presented in Main. The results are summarized in Table~\ref{tab:GP_Results}. This analysis tested the \textit{Fermi}-LAT $\mathbf{\pi^0}$, KRA$_\gamma^5$, and KRA$_\gamma^{50}$ templates. The CRINGE template was not included, as it was not used in the previous GP analysis\autocite{IceCube:2023ame}.

While the flux measurements are consistent, the differences in significance between the values listed for DNNC in Table~\ref{tab:bestfit_iceman} and that of Table~\ref{tab:GP_Results} are attributable to improved ice modelling in simulation~\autocite{tc-18-75-2024,IceCube:2023qua} and reconstruction~\autocite{IceCube:2024csv}, and the increased livetime with respect to previous GP analysis~\autocite{IceCube:2023ame}. As described in the Methods, the updated calibrations lead to more robust reconstructed energies and directions, as well as increased consistency between simulation and data.

The likelihood test finds that all three templates exceed \SI{5}{\textsigma}, indicating that the higher significance with respect to the results from \autocite{IceCube:2023ame} (shown in Table~\ref{tab:GP_Results}) is primarily driven by the addition of the last two years of data, as is expected of steady emission.

\begin{table}[h!]
\centering
\caption{\textbf{Flux/Model Norm \& Local Significance Using Extended Cascades Sample.} Summary of each template's best-fit flux/model norm and local significance tested with the 10yr shower sample from the previous GP analysis~\autocite{IceCube:2023ame} and the 12yr shower sample using the same reconstruction and processing methods as in the previous GP analysis~\autocite{IceCube:2023ame}. For the \textit{Fermi} $\pi^0$ template, fluxes are shown as $\mathrm{E}^2\frac{dN}{dE}$ at 100~TeV in units of $10^{-12}$~TeV~cm$^{-2}$~s$^{-1}$. For the KRA$_\gamma$ templates, the best-fit model normalization is shown in units of model flux, $\Phi_\mathrm{model}$. \label{tab:GP_Results}}

\begin{threeparttable}
  \begin{tabular}{l c c c}
    \toprule
    \textbf{Template} & \textbf{Livetime} & \textbf{Flux/Model Norm} &  \textbf{Local Significance} \\
    \midrule
    \multirow{2}{*}{\textit{Fermi} $\pi^0$}
      & 10yrs & \asym{21.8}{5.3}{4.9} & \SI{4.71}{\textsigma} \\
      & 12yrs\tnote{*} & \asym{22.0}{4.4}{4.3} & \SI{5.42}{\textsigma} \\
    \midrule
    \multirow{2}{*}{KRA$_\gamma^{5}$}
      & 10\,yrs & \asym{0.55}{0.18}{0.15} $\times$ $\Phi_\mathrm{model}$ & \SI{4.37}{\textsigma} \\
      & 12\,yrs\tnote{*} & \asym{0.61}{0.13}{0.12} $\times$ $\Phi_\mathrm{model}$ & \SI{5.66}{\textsigma} \\
    \midrule
    \multirow{2}{*}{KRA$_\gamma^{50}$}
      & 10\,yrs & \asym{0.37}{0.13}{0.11} $\times$ $\Phi_\mathrm{model}$ & \SI{3.96}{\textsigma} \\
      & 12\,yrs\tnote{*}& \asym{0.44}{0.10}{0.09} $\times$ $\Phi_\mathrm{model}$ & \SI{5.50}{\textsigma} \\
    \bottomrule
  \end{tabular}
  \begin{tablenotes}
    \item[*] Statistical uncertainties only
  \end{tablenotes}
\end{threeparttable}
\end{table}

\section{Galactic diffuse models and data release}

The four Galactic diffuse emission models tested in this work—\textit{Fermi}-LAT $\pi^0$, KRA$_\gamma^{5}$, KRA$_\gamma^{50}$, and CRINGE—cover a range of assumptions for cosmic-ray transport, source distributions, and high-energy emission mechanisms. 

The \textit{Fermi}-LAT $\pi^{0}$ template~\autocite{Fermi-LAT:2012edv} assumes spatially uniform CR diffusion and a fixed $E^{-2.7}$ neutrino spectrum. Its shape is broader in longitude and less centrally concentrated than the other tested templates. The KRA$_\gamma^5$ and KRA$_\gamma^{50}$ templates~\autocite{Gaggero:2015xza} assume spatially dependent diffusion and convection of CR within the Galactic disk, inferred from $\gamma$-ray data. This yields spectra in the inner Galaxy that are harder than the fixed $E^{-2.7}$ index of the Fermi–LAT $\pi^{0}$ template and a shape that is more concentrated towards the Galactic Centre. The two variants differ only in the maximum source rigidity, $R_{\max}=5~\mathrm{PV}$ for KRA$_\gamma^5$ and $R_{\max}=50~\mathrm{PV}$ for KRA$_\gamma^{50}$, leading to a correspondingly lower or higher high-energy cutoff in the predicted neutrino spectrum. A fourth, the CRINGE template~\autocite{Schwefer:2022zly}, is derived from a global fit to CR data, based on a rigidity-dependent diffusion model that leads to multiple rigidity breaks. The addition of an unresolved, galactic sources component allows for it to account for the gamma-ray emission observed by LHAASO and Tibet-AS$\gamma$, while making the neutrino spectrum harder along the GP thus resulting in a spatial distribution and spectrum that roughly falls between the $\pi^{0}$ and KRA$_\gamma$ predictions. These models have been utilized in previous GP searches~\autocite{IceCube:2017trr,IceCube:2019lzm,IceCube:2023ame,IceCube:2025zyb,IceCube:2023hou}.

All models are calibrated to reproduce cosmic-ray measurements and diffuse $\gamma$-ray emission at GeV to TeV energies, and therefore yield more similar neutrino flux predictions at lower energies, as illustrated around $\sim$10\,GeV in Extended Data Fig.~10.
Additionally shown are the resulting neutrino spectra for each model, evaluated over the inner Galactic region ($|b|\le15^\circ$, $|\ell|\le20^\circ$). Thin lines indicate the nominal model predictions, while thick lines represent the spectra scaled by the best-fit normalizations obtained from the global likelihood analysis. The IceCube data constrain the Galactic emission through a combination of spatial and energy information, as detailed in the Methods, with the spatial morphology of the inner Galaxy providing strong discriminating power.

After fitting, all models yield consistent flux levels in the inner Galactic region despite their differing high-energy predictions. In particular, the factor-of-five scaling required for the \textit{Fermi}-LAT $\pi^0$ template reflects a mismatch between its nominal prediction and the observed neutrino flux. While differences in spectral shape can contribute to this scaling, the fit is primarily driven by the overall flux level in the energy range where the detector is most sensitive. More detailed constraints on the spectral shape of the galactic neutrino flux in different regions of the GP will be presented in forthcoming IceCube analyses~\autocite{IceCube:2023hou,IceCube:2025pab}.

To facilitate reproducibility and further studies, the Galactic emission templates used in this analysis—including per-pixel flux maps as a function of energy, binning definitions, and best-fit normalizations—are publicly released as part of this publication.

\printbibliography
\end{refsection}

%% file: references.bib
@article{Ferriere:2001rg,
    author = "Ferriere, Katia M.",
    title = "{The interstellar environment of our galaxy}",
    eprint = "astro-ph/0106359",
    archivePrefix = "arXiv",
    doi = "10.1103/RevModPhys.73.1031",
    journal = "Rev. Mod. Phys.",
    volume = "73",
    pages = "1031--1066",
    year = "2001"
}

@article{Reid:2004rd,
    author = "Reid, Mark J. and Brunthaler, A.",
    title = "{The Proper motion of Sgr A*. 2. The Mass of Sgr A*}",
    eprint = "astro-ph/0408107",
    archivePrefix = "arXiv",
    doi = "10.1086/424960",
    journal = "Astrophys. J.",
    volume = "616",
    pages = "872--884",
    year = "2004"
}

@article{IceCube:2021rpz,
    author = "Aartsen, M. G. and others",
    collaboration = "IceCube",
    title = "{Detection of a particle shower at the Glashow resonance with IceCube}",
    eprint = "2110.15051",
    archivePrefix = "arXiv",
    primaryClass = "hep-ex",
    doi = "10.1038/s41586-021-03256-1",
    journal = "Nature",
    volume = "591",
    number = "7849",
    pages = "220--224",
    year = "2021",
    note = "[Erratum: Nature 592, E11 (2021)]"
}

@article{IceCube:2022der,
    author = "Abbasi, R. and others",
    collaboration = "IceCube",
    title = "{Evidence for neutrino emission from the nearby active galaxy NGC 1068}",
    eprint = "2211.09972",
    archivePrefix = "arXiv",
    primaryClass = "astro-ph.HE",
    doi = "10.1126/science.abg3395",
    journal = "Science",
    volume = "378",
    number = "6619",
    pages = "538--543",
    year = "2022",
}

@article{IceCube:2023ame,
    author = "Abbasi, R. and others",
    collaboration = "IceCube",
    title = "{Observation of high-energy neutrinos from the Galactic plane}",
    eprint = "2307.04427",
    archivePrefix = "arXiv",
    primaryClass = "astro-ph.HE",
    doi = "10.1126/science.adc9818",
    journal = "Science",
    volume = "380",
    number = "6652",
    pages = "1338--1343",
    year = "2023"
}

@article{IceCube:2013low,
    author = "Aartsen, M. G. and others",
    collaboration = "IceCube",
    title = "{Evidence for High-Energy Extraterrestrial Neutrinos at the IceCube Detector}",
    eprint = "1311.5238",
    archivePrefix = "arXiv",
    primaryClass = "astro-ph.HE",
    doi = "10.1126/science.1242856",
    journal = "Science",
    volume = "342",
    pages = "1242856",
    year = "2013"
}

@article{Glashow:1960zz,
    author = "Glashow, Sheldon L.",
    title = "{Resonant Scattering of Antineutrinos}",
    doi = "10.1103/PhysRev.118.316",
    journal = "Phys. Rev.",
    volume = "118",
    pages = "316--317",
    year = "1960"
}

@article{ParticleDataGroup:2024cfk,
    author = "Navas, S. and others",
    collaboration = "Particle Data Group",
    title = "{Review of particle physics}",
    doi = "10.1103/PhysRevD.110.030001",
    journal = "Phys. Rev. D",
    volume = "110",
    number = "3",
    pages = "030001",
    year = "2024"
}

@article{Abbasi:2021qfz,
    author = "Abbasi, R. and others",
    title = "{Improved Characterization of the Astrophysical Muon{\textendash}neutrino Flux with 9.5 Years of IceCube Data}",
    eprint = "2111.10299",
    archivePrefix = "arXiv",
    primaryClass = "astro-ph.HE",
    doi = "10.3847/1538-4357/ac4d29",
    journal = "Astrophys. J.",
    volume = "928",
    number = "1",
    pages = "50",
    year = "2022",
}

@article{IceCube:2024fxo,
    author = "Abbasi, R. and others",
    collaboration = "IceCube",
    title = "{Characterization of the astrophysical diffuse neutrino flux using starting track events in IceCube}",
    eprint = "2402.18026",
    archivePrefix = "arXiv",
    primaryClass = "astro-ph.HE",
    doi = "10.1103/PhysRevD.110.022001",
    journal = "Phys. Rev. D",
    volume = "110",
    number = "2",
    pages = "022001",
    year = "2024"
}

@Article{tc-18-75-2024,
AUTHOR = {Abbasi, R. and others},
TITLE = {In situ estimation of ice crystal properties at the South Pole using LED calibration data from the IceCube Neutrino Observatory},
JOURNAL = {The Cryosphere},
VOLUME = {18},
YEAR = {2024},
NUMBER = {1},
PAGES = {75--102},
DOI = {10.5194/tc-18-75-2024}
}

@article{IceCube:2024csv,
    author = "Abbasi, R. and others",
    collaboration = "IceCube",
    title = "{Improved modeling of in-ice particle showers for IceCube event reconstruction}",
    eprint = "2403.02470",
    archivePrefix = "arXiv",
    primaryClass = "astro-ph.HE",
    doi = "10.1088/1748-0221/19/06/P06026",
    journal = "JINST",
    volume = "19",
    number = "06",
    pages = "P06026",
    year = "2024"
}

@inproceedings{IceCube:2023gtp,
    author = "Savina, Pierpaolo and others",
    collaboration = "IceCube",
    title = "{Multi-flavour neutrino searches from the Milky Way Galaxy}",
    booktitle = "{Proc. 38th Int Cosmic Ray Conf. (ICRC2023)}",
    doi = "10.22323/1.444.1010",
    journal = "PoS",
    volume = "444",
    pages = "1010",
    year = "2023"
}

@inproceedings{Thiesmeyer:2025qgo,
    author = "Thiesmeyer, Matthias and Yuan, Tianlu and Seen, Leo and Lu, Lu and Karle, Albrecht",
    title = "{Measuring the Astrophysical Galactic Plane Neutrino Flux and Searching for Galactic PeVatrons using the IceCube Multi-Flavor Astrophysical Neutrino Sample}",
    booktitle = "{Proc. 39th International Cosmic Ray Conference (ICRC2025)}",
    eprint = "2507.08753",
    archivePrefix = "arXiv",
    primaryClass = "astro-ph.HE",
    reportNumber = "PoS-ICRC2025-1193",
    doi = "10.22323/1.501.1193",
    month = "7",
    volume = "501",
    pages = "1193",
    year = "2025"
}

@article{IceCube:2025zyb,
    author = "Abbasi, R. and others",
    collaboration = "IceCube",
    title = "{Time-integrated Southern-sky Neutrino Source Searches with 10 yr of IceCube Starting-track Events at Energies Down to 1 TeV}",
    eprint = "2501.16440",
    archivePrefix = "arXiv",
    primaryClass = "astro-ph.HE",
    doi = "10.3847/1538-4357/ae2c86",
    journal = "Astrophys. J.",
    volume = "998",
    number = "1",
    pages = "37",
    year = "2026"
}

@article{Beacom:2002vi,
    author = "Beacom, John F. and Bell, Nicole F. and Hooper, Dan and Pakvasa, Sandip and Weiler, Thomas J.",
    title = "{Decay of High-Energy Astrophysical Neutrinos}",
    eprint = "hep-ph/0211305",
    archivePrefix = "arXiv",
    reportNumber = "FERMILAB-PUB-02-243-A",
    doi = "10.1103/PhysRevLett.90.181301",
    journal = "Phys. Rev. Lett.",
    volume = "90",
    pages = "181301",
    year = "2003"
}

@article{Kobayashi:2000md,
    author = "Kobayashi, Makoto and Lim, C. S.",
    title = "{Pseudo Dirac scenario for neutrino oscillations}",
    eprint = "hep-ph/0012266",
    archivePrefix = "arXiv",
    reportNumber = "KEK-TH-733, KOBE-TH-00-10",
    doi = "10.1103/PhysRevD.64.013003",
    journal = "Phys. Rev. D",
    volume = "64",
    pages = "013003",
    year = "2001"
}

@article{Ioka:2014kca,
    author = "Ioka, Kunihto and Murase, Kohta",
    title = "{IceCube PeV{\textendash}EeV neutrinos and secret interactions of neutrinos}",
    eprint = "1404.2279",
    archivePrefix = "arXiv",
    primaryClass = "astro-ph.HE",
    reportNumber = "KEK-TH-1723, KEK-COSMO-141",
    doi = "10.1093/ptep/ptu090",
    journal = "PTEP",
    volume = "2014",
    number = "6",
    pages = "061E01",
    year = "2014"
}

@article{Ng:2014pca,
    author = "Ng, Kenny C. Y. and Beacom, John F.",
    title = "{Cosmic neutrino cascades from secret neutrino interactions}",
    eprint = "1404.2288",
    archivePrefix = "arXiv",
    primaryClass = "astro-ph.HE",
    doi = "10.1103/PhysRevD.90.065035",
    journal = "Phys. Rev. D",
    volume = "90",
    number = "6",
    pages = "065035",
    year = "2014",
    note = "[Erratum: Phys.Rev.D 90, 089904 (2014)]"
}

@article{IceCube:2016zyt,
    author = "Aartsen, M. G. and others",
    collaboration = "IceCube",
    title = "{The IceCube Neutrino Observatory: Instrumentation and Online Systems}",
    eprint = "1612.05093",
    archivePrefix = "arXiv",
    primaryClass = "astro-ph.IM",
    doi = "10.1088/1748-0221/12/03/P03012",
    journal = "JINST",
    volume = "12",
    number = "03",
    pages = "P03012",
    year = "2017",
    note = "[Erratum: JINST 19, E05001 (2024)]"
}

@article{Fermi-LAT:2012edv,
    author = "Ackermann, M. and others",
    collaboration = "Fermi-LAT",
    title = "{Fermi-LAT Observations of the Diffuse Gamma-Ray Emission: Implications for Cosmic Rays and the Interstellar Medium}",
    eprint = "1202.4039",
    archivePrefix = "arXiv",
    primaryClass = "astro-ph.HE",
    doi = "10.1088/0004-637X/750/1/3",
    journal = "Astrophys. J.",
    volume = "750",
    pages = "3",
    year = "2012"
}

@article{Gaggero:2015xza,
    author = "Gaggero, Daniele and Grasso, Dario and Marinelli, Antonio and Urbano, Alfredo and Valli, Mauro",
    title = "{The gamma-ray and neutrino sky: A consistent picture of Fermi-LAT, Milagro, and IceCube results}",
    eprint = "1504.00227",
    archivePrefix = "arXiv",
    primaryClass = "astro-ph.HE",
    doi = "10.1088/2041-8205/815/2/L25",
    journal = "Astrophys. J. Lett.",
    volume = "815",
    number = "2",
    pages = "L25",
    year = "2015"
}

@article{Schwefer:2022zly,
    author = "Schwefer, Georg and Mertsch, Philipp and Wiebusch, Christopher",
    title = "{Diffuse Emission of Galactic High-energy Neutrinos from a Global Fit of Cosmic Rays}",
    eprint = "2211.15607",
    archivePrefix = "arXiv",
    primaryClass = "astro-ph.HE",
    reportNumber = "TTK-22-40",
    doi = "10.3847/1538-4357/acc1e2",
    journal = "Astrophys. J.",
    volume = "949",
    number = "1",
    pages = "16",
    year = "2023"
}

@article{KM3NeT:2024paj,
    author = "Aiello, S. and others",
    collaboration = "KM3NeT",
    title = "{Astronomy potential of KM3NeT/ARCA}",
    eprint = "2402.08363",
    archivePrefix = "arXiv",
    primaryClass = "astro-ph.HE",
    doi = "10.1140/epjc/s10052-024-13137-2",
    journal = "Eur. Phys. J. C",
    volume = "84",
    number = "9",
    pages = "885",
    year = "2024"
}

@inproceedings{NIPS2017_6449f44a,
 author = {Ke, Guolin and Meng, Qi and Finley, Thomas and Wang, Taifeng and Chen, Wei and Ma, Weidong and Ye, Qiwei and Liu, Tie-Yan},
 booktitle = {Advances in Neural Information Processing Systems},
 editor = {I. Guyon and U. Von Luxburg and S. Bengio and H. Wallach and R. Fergus and S. Vishwanathan and R. Garnett},
 pages = {},
 publisher = {Curran Associates, Inc.},
 title = {LightGBM: A Highly Efficient Gradient Boosting Decision Tree},
 %url = {https://proceedings.neurips.cc/paper_files/paper/2017/file/6449f44a102fde848669bdd9eb6b76fa-Paper.pdf},
 volume = {30},
 year = {2017},
 keywords = {methods}
}

@inproceedings{IceCube:2025ixg,
    author = "Seen, Leo and Yuan, Tianlu and Lu, Lu and Thiesmeyer, M. and Karle, Albrecht and others",
    collaboration = "IceCube",
    title = "{Enhancing searches for astrophysical neutrino sources in IceCube with machine learning and improved spatial modeling}",
    booktitle = "{Proc. 39th Int Cosmic Ray Conf. (ICRC2025)}",
    eprint = "2507.08132",
    archivePrefix = "arXiv",
    primaryClass = "astro-ph.HE",
    reportNumber = "PoS-ICRC2025-1169",
    doi = "10.22323/1.501.1169",
    journal = "PoS",
    volume = "501",
    pages = "1169",
    year = "2025"
}

@inproceedings{IceCube:2023qua,
    author = "Abbasi, Rasha and others",
    collaboration = "IceCube",
    title = "{An improved mapping of ice layer undulations for the IceCube Neutrino Observatory}",
    booktitle = "{Proc. 38th Int Cosmic Ray Conf. (ICRC2023)}",
    eprint = "2307.13951",
    archivePrefix = "arXiv",
    primaryClass = "astro-ph.HE",
    reportNumber = "PoS-ICRC2023-975",
    doi = "10.22323/1.444.0975",
    journal = "PoS",
    volume = "444",
    pages = "975",
    year = "2023"
}

@article{IceCube:2013llx,
    author = "Aartsen, M. G. and others",
    collaboration = "IceCube",
    title = "{Measurement of South Pole ice transparency with the IceCube LED calibration system}",
    eprint = "1301.5361",
    archivePrefix = "arXiv",
    primaryClass = "astro-ph.IM",
    doi = "10.1016/j.nima.2013.01.054",
    journal = "Nucl. Instrum. Meth. A",
    volume = "711",
    pages = "73--89",
    year = "2013"
}

@article{Feldman:1997qc,
    author = "Feldman, Gary J. and Cousins, Robert D.",
    title = "{A Unified approach to the classical statistical analysis of small signals}",
    eprint = "physics/9711021",
    archivePrefix = "arXiv",
    reportNumber = "HUTP-97-A096",
    doi = "10.1103/PhysRevD.57.3873",
    journal = "Phys. Rev. D",
    volume = "57",
    pages = "3873--3889",
    year = "1998"
}

@article{Braun:2009wp,
    author = "Braun, Jim and Baker, Mike and Dumm, Jon and Finley, Chad and Karle, Albrecht and Montaruli, Teresa",
    title = "{Time-Dependent Point Source Search Methods in High Energy Neutrino Astronomy}",
    eprint = "0912.1572",
    archivePrefix = "arXiv",
    primaryClass = "astro-ph.IM",
    doi = "10.1016/j.astropartphys.2010.01.005",
    journal = "Astropart. Phys.",
    volume = "33",
    pages = "175--181",
    year = "2010",
}

@article{IceCube:2025lev,
    author = "Abbasi, R. and others",
    collaboration = "IceCube",
    title = "{All-sky Neutrino Point-source Search with IceCube Combined Track and Cascade Data}",
    eprint = "2507.07275",
    archivePrefix = "arXiv",
    primaryClass = "astro-ph.HE",
    doi = "10.3847/1538-4357/ae113f",
    journal = "Astrophys. J.",
    volume = "995",
    number = "1",
    pages = "11",
    year = "2025"
}

@article{ANTARES:2020srt,
    author = "Albert, A. and others",
    collaboration = "ANTARES, IceCube",
    title = "{ANTARES and IceCube Combined Search for Neutrino Point-like and Extended Sources in the Southern Sky}",
    eprint = "2001.04412",
    archivePrefix = "arXiv",
    primaryClass = "astro-ph.HE",
    doi = "10.3847/1538-4357/ab7afb",
    journal = "Astrophys. J.",
    volume = "892",
    pages = "92",
    year = "2020"
}

@article{IceCube:2019lzm,
    author = "Aartsen, M. G. and others",
    collaboration = "IceCube",
    title = "{Search for Sources of Astrophysical Neutrinos Using Seven Years of IceCube Cascade Events}",
    eprint = "1907.06714",
    archivePrefix = "arXiv",
    primaryClass = "astro-ph.HE",
    doi = "10.3847/1538-4357/ab4ae2",
    journal = "Astrophys. J.",
    volume = "886",
    pages = "12",
    year = "2019"
}

@article{ANTARES:2018nyb,
    author = "Albert, A. and others",
    collaboration = "ANTARES, IceCube",
    title = "{Joint Constraints on Galactic Diffuse Neutrino Emission from the ANTARES and IceCube Neutrino Telescopes}",
    eprint = "1808.03531",
    archivePrefix = "arXiv",
    primaryClass = "astro-ph.HE",
    doi = "10.3847/2041-8213/aaeecf",
    journal = "Astrophys. J. Lett.",
    volume = "868",
    number = "2",
    pages = "L20",
    year = "2018"
}

@article{IceCube:2017trr,
    author = "Aartsen, M. G. and others",
    collaboration = "IceCube",
    title = "{Constraints on Galactic Neutrino Emission with Seven Years of IceCube Data}",
    eprint = "1707.03416",
    archivePrefix = "arXiv",
    primaryClass = "astro-ph.HE",
    doi = "10.3847/1538-4357/aa8dfb",
    journal = "Astrophys. J.",
    volume = "849",
    number = "1",
    pages = "67",
    year = "2017"
}

@article{IceCube:2013cdw,
    author = "Aartsen, M. G. and others",
    collaboration = "IceCube",
    title = "{First observation of PeV-energy neutrinos with IceCube}",
    eprint = "1304.5356",
    archivePrefix = "arXiv",
    primaryClass = "astro-ph.HE",
    doi = "10.1103/PhysRevLett.111.021103",
    journal = "Phys. Rev. Lett.",
    volume = "111",
    pages = "021103",
    year = "2013"
}

@article{IceCube:2025tgp,
    author = "Abbasi, R. and others",
    collaboration = "IceCube",
    title = "{Evidence for a Spectral Break or Curvature in the Spectrum of Astrophysical Neutrinos from 5 TeV--10 PeV}",
    eprint = "2507.22233",
    archivePrefix = "arXiv",
    primaryClass = "astro-ph.HE",
    doi = "10.1103/2gh9-d4q7",
    journal = "Phys. Rev. Lett.",
    volume = "136",
    pages = "121002",
    year = "2026"
}

@article{IceCube:2015gsk,
    author = "Aartsen, M. G. and others",
    collaboration = "IceCube",
    title = "{A combined maximum-likelihood analysis of the high-energy astrophysical neutrino flux measured with IceCube}",
    eprint = "1507.03991",
    archivePrefix = "arXiv",
    primaryClass = "astro-ph.HE",
    doi = "10.1088/0004-637X/809/1/98",
    journal = "Astrophys. J.",
    volume = "809",
    number = "1",
    pages = "98",
    year = "2015"
}

@inproceedings{IceCube:2025gna,
    author = "Yildizci, Emre Burak and Rechav, Zoe and Lu, Lu and others",
    collaboration = "IceCube",
    title = "{Measurement of All Flavor PeV Neutrino Flux using Combined Datasets from IceCube}",
    booktitle = "{Proc. 39th Int Cosmic Ray Conf. (ICRC2025)}",
    eprint = "2508.05886",
    archivePrefix = "arXiv",
    primaryClass = "astro-ph.HE",
    reportNumber = "PoS-ICRC2025-1217",
    doi = "10.22323/1.501.1217",
    journal = "PoS",
    volume = "501",
    pages = "1217",
    year = "2025"
}

@article{IceCube:2019lxi,
    author = "Aartsen, M. G. and others",
    collaboration = "IceCube",
    title = "{Efficient propagation of systematic uncertainties from calibration to analysis with the SnowStorm method in IceCube}",
    eprint = "1909.01530",
    archivePrefix = "arXiv",
    primaryClass = "hep-ex",
    doi = "10.1088/1475-7516/2019/10/048",
    journal = "JCAP",
    volume = "10",
    pages = "048",
    year = "2019"
}

@article{Gaisser:2014bja,
    author = "Gaisser, Thomas K. and Jero, Kyle and Karle, Albrecht and van Santen, Jakob",
    title = "{Generalized self-veto probability for atmospheric neutrinos}",
    eprint = "1405.0525",
    archivePrefix = "arXiv",
    primaryClass = "astro-ph.HE",
    doi = "10.1103/PhysRevD.90.023009",
    journal = "Phys. Rev. D",
    volume = "90",
    number = "2",
    pages = "023009",
    year = "2014"
}

@article{Arguelles:2018awr,
    author = {Arg{\"u}elles, Carlos A. and Palomares-Ruiz, Sergio and Schneider, Austin and Wille, Logan and Yuan, Tianlu},
    title = "{Unified atmospheric neutrino passing fractions for large-scale neutrino telescopes}",
    eprint = "1805.11003",
    archivePrefix = "arXiv",
    primaryClass = "hep-ph",
    reportNumber = "IFIC/18-24, IFIC-18-24",
    doi = "10.1088/1475-7516/2018/07/047",
    journal = "JCAP",
    volume = "07",
    pages = "047",
    year = "2018"
}

@inproceedings{IceCube:2023hou,
    author = "Fuerst, Philipp Michael and others",
    collaboration = "IceCube",
    title = "{Galactic and Extragalactic Analysis of the Astrophysical Muon Neutrino Flux with 12.3 years of IceCube Track Data}",
    booktitle = "{Proc. 38th Int Cosmic Ray Conf. (ICRC2023)}",
    eprint = "2308.08233",
    archivePrefix = "arXiv",
    primaryClass = "astro-ph.HE",
    reportNumber = "PoS-ICRC2023-1046",
    doi = "10.22323/1.444.1046",
    journal = "PoS",
    volume = "444",
    pages = "1046",
    year = "2023"
}

@article{IceCube:2025pab,
    author = "Neste, Ludwig and others",
    collaboration = "IceCube",
    title = "{Measuring the Neutrino Flux in Segments along the Galactic Plane with IceCube}",
    eprint = "2507.08097",
    archivePrefix = "arXiv",
    primaryClass = "astro-ph.HE",
    reportNumber = "PoS-ICRC2025-1130",
    doi = "10.22323/1.501.1130",
    journal = "PoS",
    volume = "ICRC2025",
    pages = "1130",
    year = "2025"
}

@article{IceCube:2015rnn,
    author = "Aartsen, M. G. and others",
    collaboration = "IceCube",
    title = "{Search for Dark Matter Annihilation in the Galactic Center with IceCube-79}",
    eprint = "1505.07259",
    archivePrefix = "arXiv",
    primaryClass = "astro-ph.HE",
    doi = "10.1140/epjc/s10052-015-3713-1",
    journal = "Eur. Phys. J. C",
    volume = "75",
    number = "10",
    pages = "492",
    year = "2015"
}

@article{Gorski:2004by,
    author = "G{\'o}rski, K. M. and Hivon, E. and Banday, A. J. and Wandelt, B. D. and Hansen, F. K. and Reinecke, M. and Bartelman, M.",
    title = "{HEALPix - A Framework for high resolution discretization, and fast analysis of data distributed on the sphere}",
    eprint = "astro-ph/0409513",
    archivePrefix = "arXiv",
    doi = "10.1086/427976",
    journal = "Astrophys. J.",
    volume = "622",
    pages = "759--771",
    year = "2005"
}

@article{Ambrosone:2023hsz,
    author = "Ambrosone, Antonio and Groth, Kathrine M{\o}rch and Peretti, Enrico and Ahlers, Markus",
    title = "{Galactic diffuse neutrino emission from sources beyond the discovery horizon}",
    eprint = "2306.17285",
    archivePrefix = "arXiv",
    primaryClass = "astro-ph.HE",
    doi = "10.1103/PhysRevD.109.043007",
    journal = "Phys. Rev. D",
    volume = "109",
    number = "4",
    pages = "043007",
    year = "2024"
}

@article{Gagliardini:2024een,
    author = "Gagliardini, Silvia and Langella, Aurora and Guetta, Dafne and Capone, Antonio",
    title = "{Neutrino Fluxes from Different Classes of Galactic Sources}",
    eprint = "2403.05288",
    archivePrefix = "arXiv",
    primaryClass = "astro-ph.HE",
    doi = "10.3847/1538-4357/ad4960",
    journal = "Astrophys. J.",
    volume = "969",
    number = "2",
    pages = "161",
    year = "2024"
}

@article{Fang:2024fyd,
    author = "Fang, Ke and Halzen, Francis",
    title = "{IceCube results and perspective for neutrinos from LHAASO sources}",
    doi = "10.1016/j.jheap.2024.07.001",
    journal = "JHEAp",
    volume = "43",
    pages = "140--152",
    year = "2024"
}

@article{Sun:2023ibg,
    author = "Sun, Dong-Xu and Zhang, Pei-Pei and Yuan, Qiang and Liu, Wei and Guo, Yi-Qing",
    title = "{Multimessenger observations support cosmic ray interactions surrounding acceleration sources}",
    eprint = "2307.02372",
    archivePrefix = "arXiv",
    primaryClass = "astro-ph.HE",
    doi = "10.1103/PhysRevD.110.103039",
    journal = "Phys. Rev. D",
    volume = "110",
    number = "10",
    pages = "103039",
    year = "2024"
}

@article{Carpio:2025arz,
    author = "Carpio, Jose A. and Kheirandish, Ali and Zhang, Bing",
    title = "{Multimessenger emission from very-high-energy black hole-jet systems in the milky way}",
    eprint = "2506.22550",
    archivePrefix = "arXiv",
    primaryClass = "astro-ph.HE",
    doi = "10.1016/j.jheap.2025.100538",
    journal = "JHEAp",
    volume = "51",
    pages = "100538",
    year = "2026"
}

@article{Peretti:2024ecg,
    author = "Peretti, Enrico and Petropoulou, Maria and Vasilopoulos, Georgios and Gabici, Stefano",
    title = "{Particle acceleration and multi-messenger radiation from ultra-luminous X-ray sources - A new class of Galactic PeVatrons}",
    eprint = "2411.08762",
    archivePrefix = "arXiv",
    primaryClass = "astro-ph.HE",
    doi = "10.1051/0004-6361/202452987",
    journal = "Astron. Astrophys.",
    volume = "698",
    pages = "A188",
    year = "2025"
}

@article{Moghadam:2026jmy,
    author = "Moghadam, Mohadeseh Ozlati and Egberts, Kathrin and Batzofin, Rowan and Steppa, Constantin and Bernardini, Elisa",
    title = "{Estimating the contribution of galactic neutrino sources}",
    eprint = "2602.13815",
    archivePrefix = "arXiv",
    primaryClass = "astro-ph.HE",
    doi = "10.1016/j.astropartphys.2026.103216",
    journal = "Astropart. Phys.",
    volume = "178",
    pages = "103216",
    year = "2026"
}

@article{Menchiari:2026ney,
    author = "Menchiari, Stefano and Celli, Silvia and Vecchiotti, Vittoria and Morlino, Giovanni and Peron, Giada and L{\'o}pez-Coto, Rub{\'e}n",
    title = "{Modelling Galactic neutrino emission: contributions from massive star clusters and interstellar cosmic rays}",
    eprint = "2606.09604",
    archivePrefix = "arXiv",
    primaryClass = "astro-ph.HE",
    month = "6",
    year = "2026"
}

@article{Mancina:2021jbk,
    author = "Mancina, S. and Silva, M.",
    collaboration = "IceCube",
    title = "{Astrophysical neutrino source searches with IceCube starting track events}",
    doi = "10.1088/1748-0221/16/09/C09024",
    journal = "JINST",
    volume = "16",
    number = "09",
    pages = "C09024",
    year = "2021"
}
